\documentclass[aps,prd,showpacs,twocolumn,floatfix,nofootinbib]{revtex4-1}
\usepackage[dvipdfm]{graphicx}
\usepackage{amsmath,amssymb}
\usepackage{color,ulem}
\newcommand{\fD}{{\raise1.0ex\hbox{${}^{\,\circ}$}}\!\!\!\!D}

\begin{document}
\title{Reducing orbital eccentricity in initial data of binary neutron
stars}

\author{
Koutarou Kyutoku$^1$, Masaru Shibata$^2$, and Keisuke Taniguchi$^3$
}
\affiliation{
$^1$Department of Physics, University of Wisconsin-Milwaukee, P.O. Box
413, Milwaukee, Wisconsin 53201, USA\\
$^2$Yukawa Institute for Theoretical Physics, Kyoto University, Kyoto
606-8502, Japan\\
$^3$Graduate School of Arts and Sciences, University of Tokyo, Komaba,
Meguro, Tokyo 153-8902, Japan
}

\date{\today}

\begin{abstract}
 We develop a method to compute low-eccentricity initial data of binary
 neutron stars required to perform realistic simulations in numerical
 relativity. The orbital eccentricity is controlled by adjusting the
 orbital angular velocity of a binary and incorporating an approaching
 relative velocity of the neutron stars. These modifications improve the
 solution primarily through the hydrostatic equilibrium equation for the
 binary initial data. The orbital angular velocity and approaching
 velocity of initial data are updated iteratively by performing time
 evolutions over $\sim$ 3 orbits. We find that the eccentricity can be
 reduced by an order of magnitude compared to standard quasicircular
 initial data, specifically from $\sim$ 0.01 to $\lesssim$ 0.001, by
 three successive iterations for equal-mass binaries leaving $\sim$ 10
 orbits before the merger.
\end{abstract}

\maketitle

\section{Introduction} \label{sec:intro}

Fully general relativistic simulations of binary neutron star mergers
have been extensively performed in the past fifteen years (see
Ref.~\cite{faber_rasio2011} and reference therein for earlier
works). Prime targets of these studies are initially circular binaries
with irrotational velocity fields, because gravitational radiation
reaction circularizes most of the binaries before they enter sensitive
bands of ground-based gravitational-wave detectors \cite{peters1964} and
the viscosity of neutron-star matter is expected to be too low to
significantly increase the neutron-star spin via tidal effects
\cite{kochanek1992,bildsten_cutler1992}. In addition, the orbital period
of binary neutron stars right before the merger is much smaller than
typical rotational periods of observed neutron stars
\cite{lorimer2008}. Numerical-relativity simulations of binary neutron
stars have elucidated quantitatively the formation of remnant massive
neutron stars and/or collapse to black hole-disk systems
\cite{hotokezaka_kkmsst2013,hotokezaka_kss}, substantial mass ejection
\cite{hotokezaka_kkosst2013}, and gravitational waveforms during the
inspiral-merger-postmerger phases
\cite{hotokezaka_ks2013,read_bcfgkmrst2013}.

All the simulations of ``circular'' binary neutron stars have suffered
from unphysical orbital eccentricity. To perform numerical simulations
of circular binary neutron stars throughout inspiral, merger, and
postmerger phases, sufficiently circularized initial data are
necessary. To date, quasiequilibrium states,\footnote{In this paper, we
refer to initial data satisfying a subset of the Einstein equations
including constraints and hydrostatics as ``quasiequilibrium.''
Quasiequilibrium initial data derived under helical symmetry are called
specifically ``quasicircular,'' and those with an approaching velocity
are called ``low eccentricity.''} which are solutions to a subset of the
Einstein equations and hydrostatics under physical assumptions
\cite{gourgoulhon_gtmb2001,taniguchi_gourgoulhon2002,taniguchi_gourgoulhon2003,taniguchi_shibata2010},
have usually been adopted as initial data of numerical
simulations. Although the formulation for irrotational velocity fields
has been developed to a satisfactory level
\cite{bonazzola_gm1997,asada1998,shibata1998,teukolsky1998}, it has
failed to give sufficiently circularized binaries and instead has
resulted in eccentricities $e \gtrsim 0.01$. Computations of
quasicircular binary neutron stars have been carried out assuming the
existence of a helical Killing vector field with the orbital angular
velocity $\Omega$. Because this formulation neglects the gravitational
radiation reaction and does not appropriately incorporate the
approaching velocity, a binary prepared in this manner evolves as a
slightly but appreciably eccentric binary once a numerical simulation is
launched. Some effort has been made to remove the eccentricity by adding
a post-Newton-inspired approaching velocity to quasicircular initial
data \cite{kiuchi_sst2009} or by constraint-violating superposition of
Lorentz-boosted stationary stars \cite{tsatsin_marronetti2013}, but the
eccentricity is not reduced below $e \sim 0.01$.

Reducing the orbital eccentricity is an urgent task in numerical
relativity, and the most important reason for this is high demand to
derive accurate gravitational waveforms. Although mismatch due to the
orbital eccentricity in quasicircular initial data is reported to be
merely $\sim$ 1\% in binary black hole simulations if we focus only on
the numerical-relativity waveforms \cite{pfeiffer_bklls2007}, the
eccentricity complicates comparisons between gravitational waveforms
obtained by numerical simulations and those derived by analytic methods
both for binary black holes
\cite{buonanno_cp2007,boyle_bkmpsct2007,hannam_hgsb2008} and binary
neutron stars
\cite{baiotti_dgnr2011,bernuzzi_ntb2012,hotokezaka_ks2013}. The
eccentricity also affects hybridization of analytic and numerical
waveforms, which is necessary to create phenomenological templates
covering a wide frequency domain (see
Refs.~\cite{lackey_ksbf2012,pannarale_bks2013,foucart_ddkmopsst2013,lackey_ksbf2014}
for relevant works of black hole-neutron star binaries). The extraction
of tidal deformability from gravitational waves is fairly sensitive to
the accuracy of templates
\cite{favata2014,yagi_yunes2014,wade_colflr2014}, and thus reducing the
orbital eccentricity is critical to maximizing the precision with which
we can extract neutron-star parameters and constrain the neutron-star
equations of state from gravitational-wave observations.

In this paper, we describe a method to reduce orbital eccentricity in
initial data of binary neutron stars. The basic idea is similar to that
of eccentricity reduction for binary black holes
\cite{pfeiffer_bklls2007}. Namely, we correct initial data using their
orbital evolution obtained by dynamical simulations and repeat this
procedure until a desired value of the orbital eccentricity is
achieved. Specifically, we adjust the orbital angular velocity and
incorporate an approaching velocity. The primary difference from the
binary black hole cases resides in the method for computing initial data
of binary neutron stars with an approaching velocity. In binary black
hole problems solved in the extended conformal thin-sandwich formulation
(see Sec.~\ref{sec:initial}), the approaching velocity and eccentricity
of initial data are controlled via inner boundary conditions imposed at
black hole horizons \cite{pfeiffer_bklls2007}. Instead, we control them
by modifying hydrostatics of neutron stars. As a first attempt to reduce
the orbital eccentricity in binary neutron stars,\footnote{We noticed
that the Simulating eXtreme Spacetimes (SXS) collaboration also has
succeeded independently in reducing the orbital eccentricity when this
study was nearly completed \cite{haas}.} we aim for eccentricity lower
than $\sim 0.001$, which satisfies the requirement $e \lesssim 0.002$ of
the Numerical-Relativity and Analytical-Relativity Collaboration for
binary black holes \cite{hinder_etal2014}.

This paper is organized as follows. In Sec.~\ref{sec:initial}, we
describe our method to compute low-eccentricity initial data of binary
neutron stars with an approaching velocity. A procedure to analyze
orbital evolution is presented in Sec.~\ref{sec:simulation} with a brief
description of our simulation code, including an update from the
Baumgarte--Shapiro--Shibata--Nakamura (BSSN) formulation
\cite{shibata_nakamura1995,baumgarte_shapiro1998} to conformally
decomposed Z4 (Z4c) formulation \cite{bernuzzi_hilditch2010}. Actual
eccentricity reduction and results obtained with low-eccentricity
initial data are demonstrated in Sec.~\ref{sec:result}. Section
\ref{sec:summary} is devoted to a summary and discussions.

Greek and Latin indices denote the spacetime and space components,
respectively. Geometrical units in which $G=c=1$, where $G$ and $c$ are
the gravitational constant and speed of light, respectively, are adopted
throughout this paper.

\section{Initial data computation} \label{sec:initial}

We compute initial data of binary neutron stars in the extended
conformal thin-sandwich formulation
\cite{york1999,pfeiffer_york2003}. We also solve equations of
hydrostatics to obtain quasiequilibrium fluid configurations. The
formulation is formally very similar to that for quasicircular initial
data, for which the details are found in
Refs.~\cite{gourgoulhon_gtmb2001,taniguchi_gourgoulhon2002,taniguchi_gourgoulhon2003,taniguchi_shibata2010}. All
the numerical computations of initial data are performed with a public
multidomain spectral method library, {\small LORENE} \cite{LORENE}, and
numerical details are found in
Refs.~\cite{bonazzola_gm1998,gourgoulhon_gtmb2001}.

\subsection{Gravitational field equations} \label{ssec:grav}

Physically valid initial data have to satisfy the Hamiltonian and
momentum constraints at the very least, and some quasiequilibrium
assumptions are desired to be met for astrophysically realistic binary
initial data. In this study, we compute initial data of the induced
metric $\gamma_{ij}$ and extrinsic curvature $K_{ij}$ in the extended
conformal thin-sandwich formulation \cite{york1999,pfeiffer_york2003}. A
conformal transformation is defined by
\begin{align}
 \gamma_{ij} & = \psi^4 \hat{\gamma}_{ij} \; , \; \gamma^{ij} =
 \psi^{-4} \hat{\gamma}_{ij} , \\
 A^{ij} & = \psi^{-10} \hat{A}^{ij} \; , \; A_{ij} = \psi^{-2}
 \hat{A}_{ij} ,
\end{align}
where $A_{ij}$ is the traceless part of the extrinsic curvature given as
\begin{equation}
 A_{ij} \equiv K_{ij} - \frac{1}{3} K \gamma_{ij} \; , \; K \equiv
  \gamma^{ij} K_{ij} .
\end{equation}
We handle a weighted lapse function $\Phi \equiv \alpha \psi$ instead of
the lapse function $\alpha$ itself in the computation of initial data
for the sake of numerical accuracy, whereas the shift vector $\beta^i$
is handled as it is. A traceless evolution tensor of the conformal
induced metric,
\begin{equation}
 \hat{u}_{ij} = \partial_t \hat{\gamma}_{ij} ,
\end{equation}
with $\hat{\gamma}^{ij} \hat{u}_{ij} = 0$ is also introduced as freely
specifiable data in this formulation.

To obtain a quasiequilibrium configuration of binary neutron stars, we
impose conditions,
\begin{equation}
 \hat{\gamma}_{ij} = f_{ij} \; , \; K = 0 \; , \; \hat{u}_{ij} = 0 \; ,
  \; \partial_t K = 0 , \label{eq:quasieq}
\end{equation}
on freely specifiable data. Here, $f_{ij}$ is the flat 3-metric. In
principle, attention has to be paid for the frame in which equations are
solved, since the latter two stationarity conditions could be imposed
even approximately only when the time direction is chosen to agree with
the binary motion. Specifically, the change of frames amounts to
adopting different shift vectors, which appears in various terms of
gravitational field equations. It turns out that, however, the equations
are unaffected by the addition of a reasonable approaching velocity as
pointed out in Ref.~\cite{pfeiffer_bklls2007}. The reason for this is
that the difference of the shift vector enters the equations only
through (i) the conformal Killing operator associated with
$\hat{\gamma}_{ij}$ and (ii) the Lie derivative of $K$. This is the very
reason why a rotational shift vector of the form $\Omega \left(
\partial_\varphi \right)^\mu$ has not appeared explicitly in the
computation of quasicircular initial data
\cite{gourgoulhon_gtmb2001,taniguchi_gourgoulhon2002,taniguchi_gourgoulhon2003,taniguchi_shibata2010}.

One plausible way to incorporate an approaching velocity may be to add
uniform contraction to the helical Killing vector in the manner
\cite{pfeiffer_bklls2007}
\begin{equation}
 \xi^\mu = \left( \partial_t \right)^\mu + \Omega \left(
	    \partial_\varphi \right)^\mu + v \frac{r}{r_0} \left(
	    \partial_r \right)^\mu \label{eq:radial}
\end{equation}
and impose the quasiequilibrium conditions, Eq.~\eqref{eq:quasieq}, in
the time direction given by $\xi^\mu$ [see also
Eq.~\eqref{eq:trans}]. Here, $v$ (negative for the approaching velocity)
and $r_0$ should be regarded as the radial velocity and separation from
the coordinate origin, respectively, averaged over binary
members.\footnote{Alternatively, $v$ and $r_0$ may be regarded as the
radial velocity and separation, respectively, of the binary. In this
case, the value of $v$ in this paper has to be replaced by $2v$.}  For
an equal-mass binary, $v$ and $r_0$ should be common in each member, and
the averaging is not required. The radial velocity term $r ( \partial_r
)^i$ is a conformal Killing vector of the flat metric, and hence this
term does not affect gravitational field equations. The boundary
condition of the shift vector (see below) is also unaffected, because a
conformal Killing vector is a homogeneous solution of the vectorial
Laplacian.

The equations to be solved for gravitational fields become
\cite{taniguchi_shibata2010}
\begin{align}
 \fD^2 \psi & = - \frac{1}{8} \psi^{-7} \hat{A}_{ij} \hat{A}^{ij} - 2
 \pi \psi^5 \rho_\mathrm{H} , \label{eq:xctsconf} \\
 \fD^2 \beta^i + \frac{1}{3} \fD^i \fD_j \beta^j & = 2 \hat{A}^{ij}
 \fD_j \left( \Phi \psi^{-7} \right) + 16 \pi \Phi \psi^3 j^i ,
 \label{eq:xctsshift} \\
 \fD^2 \Phi & = \frac{7}{8} \Phi \psi^{-8} \hat{A}_{ij} \hat{A}^{ij} + 2
 \pi \Phi \psi^4 \left( \rho_\mathrm{H} + 2S \right) ,
 \label{eq:xctslapse} \\
 \hat{A}^{ij} & = \frac{\psi^7}{2 \Phi} \left( \fD^i \beta^j + \fD^j
 \beta^i - \frac{2}{3} f^{ij} \fD_k \beta^k \right) \label{eq:xctsextr}
 ,
\end{align}
where $\fD_i$ is the covariant derivative associated with the flat
metric, $f_{ij}$. The matter source terms are defined from the 3+1
decomposition of the energy-momentum tensor $T_{\mu \nu}$ in terms of a
future-directed unit normal vector $n^\mu$ to the constant-time
hypersurface as
\begin{align}
 \rho_\mathrm{H} & = T_{\mu \nu} n^\mu n^\nu , \\
 j_i & = - \gamma_{i\mu} T^{\mu \nu} n_\nu , \\
 S_{ij} & = \gamma_{i\mu} \gamma_{j\nu} T^{\mu \nu} .
\end{align}
The scalar elliptic equations are solved with boundary conditions,
\begin{equation}
 \psi , \Phi \to 1 \; ( r \to \infty ) , \label{eq:bcscalar}
\end{equation}
derived from the asymptotic flatness. The boundary condition on the
shift vector determines the frame in which the equations are solved, and
we can simply set
\begin{equation}
 \beta^i \to 0 \; ( r \to \infty ) , \label{eq:bcshift}
\end{equation}
according to the discussion above, remembering that $\beta^i$ obtained
with this is that for an asymptotically inertial frame.

\subsection{Hydrostatics} \label{ssec:hydro}

The neutron-star matter is modeled by a perfect fluid with zero
temperature, for which all the thermodynamic quantities are given as
functions of only one representative, in our computation of initial
data. The energy-momentum tensor takes the form
\begin{equation}
 T_{\mu \nu} = \rho h u_\mu u_\nu + P g_{\mu \nu} ,
\end{equation}
where $\rho$, $P$, $h$, and $u^\mu$ are the rest-mass density, pressure,
specific enthalpy, and 4-velocity of the fluid, respectively. The
specific enthalpy is defined by $h = 1 + \varepsilon + P / \rho$, where
$\varepsilon$ is the specific internal energy. The hydrodynamic
equations comprise the continuity equation
\begin{equation}
 \nabla_\mu \left( \rho u^\mu \right) = 0 \label{eq:cont}
\end{equation}
and the local energy-momentum conservation equation
\begin{equation}
 \nabla_\nu T^{\mu \nu} = 0 , \label{eq:emcons}
\end{equation}
where $\nabla_\mu$ is the covariant derivative associated with the
spacetime metric. Because we focus only on the irrotational velocity
field, a velocity potential $\Psi$ such that
\begin{equation}
 \nabla_\mu \Psi = h u_\mu
\end{equation}
can be introduced \cite{shibata1998,teukolsky1998}. The time component
of Eq.~\eqref{eq:emcons} in the comoving frame of the fluid merely gives
Eq.~\eqref{eq:cont} for a zero-temperature perfect fluid, and the
spatial components of Eq.~\eqref{eq:emcons}, or the relativistic Euler
equation, are shown to be integrable for an irrotational flow in the
presence of symmetry \cite{shibata1998,teukolsky1998}. A nontrivial
equation to be solved is only the continuity, Eq.~\eqref{eq:cont}, and
it is reformulated to a Poisson-like equation for $\Psi$.

A hydrostatic equilibrium can be obtained when a symmetry for
hydrodynamical fields exists. The symmetry naturally arises in a
spacetime equipped with a Killing vector field such as the helical
Killing vector. We do not assume, however, the existence of Killing
vector fields, because the approaching velocity is not compatible with
the spacetime symmetry. Instead, we only assume that all the
hydrodynamical fields are conserved when they are Lie dragged along a
symmetry vector field $\xi^\mu$.

To compute low-eccentricity initial data, a symmetry expressed by a
modified vector field, Eq.~\eqref{eq:radial}, may be desired symmetry to
be imposed in the presence of an approaching velocity. A potential
caveat with Eq.~\eqref{eq:radial}, however, is the fact that the radial
velocity term, $r ( \partial_r )^i$, is not divergence free. This
suggests that a neutron star with the approaching velocity represented
by this form has a contracting density profile. Such initial data might
introduce unphysical oscillations of neutron stars. This problem does
not arise in computations of quasicircular initial data with the helical
Killing symmetry, which is represented by a divergence-free vector
field.

This undesired possibility may be avoided by adopting a different
symmetry vector for the hydrodynamical fields from that for the
gravitational fields. Specifically, we can adopt a divergence-free
vector,
\begin{equation}
 \xi^\mu = \left( \partial_t \right)^\mu + \Omega \left(
	    \partial_\varphi \right)^\mu + v_\pm \left( \partial_x
						 \right)^\mu ,
	    \label{eq:trans}
\end{equation}
where neutron stars are assumed to lie on the $x$ axis and $v_+$ and
$v_-$ apply to neutron stars at $x>0$ and $x<0$, respectively. The
values of $v_\pm$ should be chosen to satisfy $v_+ - v_- = 2v$ [see
Eq.~\eqref{eq:radial}], and the partition may be done according to their
locations relative to the rotational axis or masses in isolation. We
should set $v_+ = - v_- = v<0$ for an approaching equal-mass binary. A
potential caveat with Eq.~\eqref{eq:trans} is inconsistency of the
symmetry for gravitational and hydrodynamical fields, which might induce
undesired behavior such as, again, neutron star oscillations once
initial data are evolved in time. Although it might seem that
Eq.~\eqref{eq:trans} can also be applied to gravitational fields
considering that the translation is isometry of the flat metric,
incorporating translational velocities $v_+<0$ in the $+x$ region and
$v_->0$ in the $-x$ region simultaneously (say) would lead to singular
behavior at the $x=0$ plane.

From our experience, no noticeable difference is found between results
obtained with these two symmetry vectors, including the oscillation of
neutron stars during time evolution, for an equal-mass
binary. Specifically, the amount of excited oscillations seems to be the
same as that for quasicircular initial data with either choice. This is
natural because the difference between the two vectors is a tiny
correction to the approaching velocity, which is small in itself
compared to the orbital velocity. Thus, the choice is a matter of taste
for an equal-mass binary. We still prefer to adopt Eq.~\eqref{eq:trans},
because this symmetry vector will give us flexibility to adjust the
partition of the approaching velocity for both binary members. This may
be useful in computing low-eccentricity initial data of unequal-mass
binaries. All the results shown in this paper are obtained with
Eq.~\eqref{eq:trans}, and results obtained with Eq.~\eqref{eq:radial}
are essentially the same as far as equal-mass binaries are concerned.

We formulate the hydrostatic equations following
Ref.~\cite{gourgoulhon_gtmb2001}. An observer with 4-velocity $v^\mu$
parallel to $\xi^\mu$ is introduced, and $v^\mu$ is decomposed in a 3+1
manner as
\begin{equation}
 v^\mu = \Gamma_0 \left( n^\mu + V^\mu \right) ,
\end{equation}
where $n_\mu V^\mu = 0$. The normalization of the 4-velocity implies
\begin{align}
 V^i & = \frac{1}{\alpha} \left[ \beta^i + \Omega \left(
 \partial_\varphi \right)^i + v_\pm \left( \partial_x \right)^i	\right]
 , \\
\Gamma_0 & = \frac{1}{\sqrt{1 - V^i V_i}} ,
\end{align}
and thus these quantities are written purely in terms of the geometric
quantities. The first integral of the relativistic Euler equation is
given by combining the irrotationality condition and assumed symmetry,
$\pounds_\xi ( h u_\mu ) = 0$. Using the expression $\xi^\mu = ( \alpha
/ \Gamma_0 ) v^\mu$, we obtain
\begin{equation}
 h \alpha \frac{\Gamma}{\Gamma_0} = C , \label{eq:firstinteg}
\end{equation}
where $\Gamma \equiv - v_\mu u^\mu$ and $C$ is a constant. This equation
should be considered as an equation to determine the specific
enthalpy. The information of the symmetry, $\xi^\mu$, is encoded in
$\Gamma_0$ (and partly in $\Gamma$) in this equation. The continuity
equation is written as an elliptic equation for $\Psi$ as
\begin{align}
 & \rho D^2 \Psi + \left( D^i \Psi \right) \left( D_i \rho \right)
 \notag \\
 = & \Gamma_n \rho h K + \Gamma_n h V^i D_i \rho \notag \\
 + & \rho \left[ \left( D^i \Psi \right) \left\{ D_i \ln \left(
 \frac{h}{\alpha} \right) \right\} + h V^i D_i \Gamma_n \right] ,
\end{align}
where $D_i$ is the covariant derivative associated with $\gamma_{ij}$
and $\Gamma_n \equiv - n_\mu u^\mu$. The key to deriving this equation
is the use of symmetry with respect to $\xi^\mu$ (not limited to a
Killing vector) in the form
\begin{equation}
 n^\mu \nabla_\mu s = - V^i D_i s
\end{equation}
for a given scalar field $s$ except for the velocity potential
\cite{teukolsky1998}. The velocity potential, $\Psi$, gives $u^\mu$, and
hence it determines $\Gamma_n$ and $\Gamma$ in a self-consistent
manner. In actual computations of binary neutron stars, we impose the
spatial conformal flatness and maximal slicing.

\subsection{Free parameters} \label{ssec:free}

The equations for gravitational and hydrodynamical fields contain free
parameters to be specified to satisfy physical requirements. We describe
our strategy for fixing these parameters to compute quasicircular and
low-eccentricity initial data separately.

When we compute quasicircular initial data, our aim is to pick a
snapshot out of a quasiequilibrium sequence of binary neutron stars with
fixed baryon rest masses and $v=0$. Hence, the coordinate separation of
the binary is essentially freely chosen to determine a particular
snapshot. This uniquely fixes the location of each binary member
relative to the rotational axis for an equal-mass binary due to its
symmetry. For an unequal-mass binary, the relative locations of each
neutron star with respect to the rotational axis have to be specified,
and one successful method is to require that the total linear momentum
vanishes \cite{taniguchi_shibata2010}. The orbital angular velocity for
a given orbital separation is determined by requiring force balance at
the stellar center \cite{gourgoulhon_gtmb2001} as
\begin{equation}
 \left. \partial_x h \right|_\mathrm{center} = 0 .
\end{equation}
Because $\Gamma_0$ is quadratic in $\Omega$, differentiation of
Eq.~\eqref{eq:firstinteg} with respect to $x$ gives a required value of
$\Omega$ for the condition above to be satisfied. Finally, the constant,
$C$, appearing in the first integral Eq.~\eqref{eq:firstinteg} is
determined so that the baryon rest mass of a neutron star takes a
desired value. For an unequal-mass binary, values of $C$ should be
determined independently for each member.

We compute low-eccentricity initial data by modifying the orbital
angular velocity, $\Omega$, and approaching velocity, $v$, of given
binary initial data. Thus, the coordinate separation between the two
maxima of the specific enthalpy is unchanged from the original initial
data, and the orbital angular velocity and approaching velocity are
specified simultaneously. The method to determine appropriate values of
them are described in the next section. The relative locations of
neutron stars should be changed in the computation of low-eccentricity
initial data to obtain a vanishing total linear momentum, whereas this
is not necessary for equal-mass binaries. The condition that determines
$C$ is the same as in the computation of quasicircular initial data.

\section{Iterative correction to initial data} \label{sec:simulation}

The orbital evolution of particular binary initial data is investigated
by dynamical simulations. In this study, we perform simulations with an
adaptive-mesh-refinement code, {\small SACRA}
\cite{yamamoto_st2008}. Corrections to the orbital angular velocity and
approaching velocity are determined by estimating the contribution of
residual eccentricity through fitting of the orbital evolution by an
analytic function.

\subsection{Time evolution} \label{ssec:sacra}

In the latest version of {\small SACRA}, the Einstein evolution
equations are solved in the Z4c formulation
\cite{bernuzzi_hilditch2010}. The formal differences from the BSSN
formulation are the addition of a new evolution variable $\Theta$, which
serves to wash out constraint violation, and the introduction of
constraint damping parameters $\kappa_1$ and $\kappa_2$. We choose
$\kappa_1 \approx 0.015 / m_0$ and $\kappa_2 = 0$, where $m_0$ is the
total mass of the binary at infinite separation, in this
study. Variables evolved similarly to the BSSN formulation are the
conformal-factor variable $W \equiv \gamma^{-1/6}$, conformal metric
$\tilde{\gamma}_{ij} \equiv \gamma^{-1/3} \gamma_{ij}$, conformally
weighted traceless part of the extrinsic curvature $\tilde{A}_{ij}
\equiv \gamma^{-1/3} A_{ij}$, modified extrinsic curvature trace
$\hat{K} \equiv K - 2 \Theta$, and so-called conformal connection
function $\tilde{\Gamma}^i$ (see Ref.~\cite{bernuzzi_hilditch2010} for
the definition in the Z4c formulation). Their evolution equations differ
from those in the BSSN formulation \cite{yamamoto_st2008} only by
modification terms given in Eq.~(3) for $\hat{K}$ and Eq.~(5) for
$\tilde{\Gamma}^i$ of Ref.~\cite{hilditch_btctb2013}. The new variable
$\Theta$ is evolved according to
\begin{align}
 \left( \partial_t - \beta^i \partial_i \right) \Theta & = \frac{1}{2}
 \alpha \left[ R - \tilde{A}_{ij} \tilde{A}^{ij} + \frac{2}{3} K^2 - 16
 \pi \rho_\mathrm{H} \right] \notag \\
 & - \alpha \kappa_1 ( 2 + \kappa_2 ) \Theta , \label{eq:evoltheta}
\end{align}
where $R$ is the scalar curvature of $\gamma_{ij}$ and $K$ in this
expression should be understood as $\hat{K} + 2 \Theta$ in the Z4c
formulation. We also add Kreiss--Oliger dissipation of the form $0.5
\times 2^{-6} ( \Delta x )^6 ( \partial_x^6 + \partial_y^6 +
\partial_z^6 )$ for all the gravitational field variables at each
intermediate Runge--Kutta time step, where $\Delta x$ is the grid
separation. Hydrodynamical evolution equations are solved in exactly the
same manner as that in previous work using {\small SACRA}.

Differences of our specific implementation from
Ref.~\cite{hilditch_btctb2013} are as follows. First, a variable related
to the conformal factor is chosen to be $W$ instead of $\chi =
W^2$. Second, the gauge variables are evolved by $K$-driver (1+log
slicing) and $\Gamma$-driver conditions of the form
\begin{align}
 \left( \partial_t - \beta^j \partial_j \right) \alpha & = - 2 \alpha
 \hat{K} , \\
 \left( \partial_t - \beta^j \partial_j \right) \beta^i & = \frac{3}{4}
 B^i , \\
 \left( \partial_t - \beta^j \partial_j \right) B^i & = \left(
 \partial_t - \beta^j \partial_j \right) \tilde{\Gamma}^i - \eta_s B^i ,
\end{align}
where $B^i$ is an auxiliary variable and $\eta_s$ is a free
parameter. We typically choose $\eta_s \approx 1 / m_0$. It should be
cautioned that different values of $\eta_s$ excite higher harmonic modes
in the coordinate orbital evolution differently \cite{purrer_hh2012},
and thus an inappropriate choice of $\eta_s$ may be problematic for the
eccentricity reduction. Finally, because the outer boundary of {\small
SACRA} has a nonsmooth rectangular shape surrounding $(x,y,z) \in [-L:L]
\times [-L:L] \times [0:L]$ with equatorial symmetry imposed at $z=0$,
we adopt simple outgoing-wave boundary conditions
\cite{shibata_nakamura1995,yamamoto_st2008} rather than
constraint-preserving and incoming-radiation-controlling ones described
in Ref.~\cite{hilditch_btctb2013}, which require a normal vector to the
boundary. To suppress unphysical incoming modes from the boundary, we
instead force the right-hand side of Eq.~\eqref{eq:evoltheta} to damp
exponentially by multiplying $\exp [ - r^2 / (L/2)^2 ]$. The same factor
is also multiplied for all $\kappa_1$ and for $\Theta$ in the source
term of the evolution equation for $\hat{K}$. This prescription is
justified, because all the modified terms vanish for physical
solutions. A potential caveat could be the breakdown of numerical
simulations due to the modification of principal parts of the evolution
equation system, but we found no onset of instability. This prescription
significantly improves the conservation of the Arnowitt--Deser--Misner
mass and total angular momentum during simulations, whereas the orbital
evolution and gravitational waves depend only weakly on this
modification.

Initial values of the lapse function and shift vector can be freely
chosen without violating the Einstein constraint equations. Instead of
using the data obtained in the extended conformal thin-sandwich
formulation, we give as initial data of the gauge variables $\alpha =
W$, $\beta^i=0$, and $B^i=0$. This choice tends to suppress unphysical
oscillations of coordinate orbital evolution associated with gauge
dynamics without eliminating modulations due to the orbital
eccentricity. By contrast, if we use $\beta^i$ obtained in the extended
conformal thin-sandwich formulation with $B^i=0$, additional
oscillations are excited in the coordinate orbital evolution with
approximately twice the frequency of eccentricity-driven
oscillations. We discuss this issue later again in
Sec.~\ref{ssec:evol}. On another front, the dependence of the coordinate
orbital evolution on the initial choice of $\alpha$ is very weak, and
our choice may be a matter of taste for binary neutron
stars.\footnote{The modification of the lapse function is essential for
black hole-neutron star binaries to ensure its positivity if initial
data are computed in the puncture framework \cite{kyutoku_st2009}.}
Initial values of $\Theta$ and $\tilde{\Gamma}^i$ are always given as
zero.

\subsection{Finding orbital evolution} \label{ssec:findorb}

The first task is to determine the orbital evolution from grid-based
dynamical simulations. We define the location of a stellar center
$x_\mathrm{NS}^i = (x_\mathrm{NS} , y_\mathrm{NS} , 0)$ at each time
step by a point of the maximum conserved rest-mass density $\rho_* =
\rho \alpha u^t \sqrt{\gamma}$ on computational grids, where two
distinct points corresponding to two stars are tracked during the
inspiral. The coordinate orbital separation $d (t)$ and orbital phase
$\phi (t)$ are directly computed as
\begin{align}
 d (t) & = \sqrt{ \left( x_\mathrm{NS,1} - x_\mathrm{NS,2} \right)^2 +
 \left( y_\mathrm{NS,1} - y_\mathrm{NS,2} \right)^2 } , \\
 \phi (t) & = \arctan \left( \frac{y_\mathrm{NS,1} -
 y_\mathrm{NS,2}}{x_\mathrm{NS,1} - x_\mathrm{NS,2}} \right) + 2 \pi N ,
\end{align}
where the label 1 or 2 stands for a member of the binary. The orbital
phase has a freedom of adding an integer multiple of $2 \pi$, and it is
fixed to be continuous by appropriately choosing an integer $N$, which
is essentially the number of orbits. The time derivative of the orbital
separation $\dot{d} (t)$, orbital angular velocity $\Omega (t) \equiv
\dot{\phi} (t)$, and its time derivative $\dot{\Omega} (t)$ are computed
from them by fourth-order finite differentiation. In this study, we use
$\dot{\Omega} (t)$ for the orbital analysis, because this should vanish
in the absence of radiation reaction for a genuinely circular orbit. In
other words, $\dot{\Omega} (t)$ should be expressed as a sum of secular
(increase) terms due to the radiation reaction and modulation terms due
to the orbital eccentricity. This should also be true of $\dot{d} (t)$,
but we do not use $\dot{d} (t)$ in this study.\footnote{This is partly
motivated by future extension to precessing black hole-neutron star
binaries \cite{buonanno_kmpt2011}.}

Care must be taken in differentiating the numerical data. Because our
numerical grids are discrete, the orbital evolution determined in this
manner inevitably contains a spurious high-frequency oscillation of the
order of the grid separation. If the time derivatives are computed
directly from raw data, this oscillation easily corrupts the result. In
this study, we smooth the orbital evolution by averaging over $\sim$
100--200 time steps before taking the time derivatives. This averaging
time interval typically amounts to 0.5 ms for simulations presented
later in Sec.~\ref{sec:result}, where exact values change within a
factor of 2 due to different numbers of time steps used for the
averaging and/or different grid resolutions. This is sufficient for the
purpose of this study, i.e., to reduce the orbital eccentricity, as far
as the eccentricity-driven modulation is largely untouched. Were we to
analyze the orbital evolution itself more carefully, a more
sophisticated estimate of the neutron star location would be needed.

After completing this work, we considered as an alternate definition of
the coordinate center of each neutron star the integral form
\begin{equation}
 x_\mathrm{NS,integ}^i \equiv \frac{\int \rho_* x^i d^3x}{\int \rho_*
  d^3 x} . \label{eq:nsinteg}
\end{equation}
This determination is found to be fully consistent with
$x_\mathrm{NS}^i$ described above and is less subject to the
high-frequency oscillations due to the discrete grids. Although the
second time derivative of the phase, $\dot{\Omega} (t)$, is still
susceptible to numerical noise, using this integration reduces the need
to average and may enhance the efficiency of the eccentricity
reduction. In addition, the orbital evolution itself can be analyzed
more reliably. The first time derivative may be further improved if it
is determined by an integral form,
\begin{equation}
 \dot{x}^i_\mathrm{NS,integ} \equiv \frac{\int \rho_* \dot{x}^i d^3
  x}{\int \rho_* d^3 x} ,
\end{equation}
where $\dot{x}^i = u^i / u^t$, using the continuity equation.

\subsection{Estimating appropriate corrections} \label{ssec:estcorr}

Information at the initial instant $t=0$ is obtained through fitting the
orbital evolution by an analytic function and is used to estimate
appropriate corrections to the orbital angular velocity $\delta \Omega$
and approaching velocity $\delta v$ of initial data. We assume that
$\dot{\Omega} (t)$ is modeled by
\begin{equation}
 \dot{\Omega} (t) = A_0 + A_1 t + B \cos ( \omega t + \phi_0 ) ,
\end{equation}
where $\{ A_0 , A_1 , B , \omega , \phi_0 \}$ are fitting parameters,
following
Refs.~\cite{pfeiffer_bklls2007,boyle_bkmpsct2007,buonanno_kmpt2011}. The
numerical evolution data are fitted to determine these parameters. The
modulation term $B \cos ( \omega t + \phi_0 )$ should be ascribed to the
eccentricity, whereas the secular terms $A_0 + A_1 t$ should result from
gravitational radiation reaction. A perfectly circularized orbit should
give $B=0$ as far as the secular terms capture evolution due to the
radiation reaction.

A time interval has to be chosen carefully to perform the
fitting. First, the duration has to be long enough to include more than
one modulation cycle associated with the eccentricity. Next, it has to
be short enough to avoid strong influence of long-term secular
evolution. Finally, some initial portion of the orbital evolution has to
be excluded from the analysis, because the evolution is significantly
affected by relaxation of initial data from the spatial conformal
flatness and of the gauge variables. Taking these issues and our
experiments into account, we use $t \in [0.5P:3P]$ for the fitting,
where $P \equiv 2 \pi / \Omega$ is the initial orbital period and
$\Omega$ is the orbital angular velocity specified in the initial value
problem. The interval spanning $\sim$ 2.5 orbits is longer than those
typically adopted in iterative eccentricity reduction of binary black
holes \cite{pfeiffer_bklls2007,boyle_bkmpsct2007,buonanno_kmpt2011}. We
expect that it is made shorter (possibly with increasing the
eccentricity reduction efficiency) by more elaborated fitting like the
one using Eq.~\eqref{eq:nsinteg}, and such an optimization is left for
the future study.

The eccentricity is estimated from the modulation term using the
knowledge of Newtonian two-body dynamics with $e \ll 1$. A variety of
proposed eccentricity estimators is summarized in
Ref.~\cite{mroue_pkt2010}, and conceptual differences among them are
analyzed in Ref.~\cite{purrer_hh2012}. We estimate the eccentricity of
initial data mainly from the fitting parameters as
\begin{equation}
 e \approx \frac{|B|}{2 \omega \Omega} , \label{eq:ecc}
\end{equation}
which is derived by an expected Newtonian relation $\Omega (t) \approx
\Omega [ 1 + 2e \sin ( \omega t + \phi_0 )]$ at $e \ll 1$. Here, the
difference between radial frequency $\omega$ and angular frequency
$\Omega$ is ascribed to the periastron advance of general relativistic
origin, and thus the distinction of these two may be arbitrary to some
extent in the Newtonian discussion. We also compute another eccentricity
estimator proposed in the literature for the consistency check in
Sec.~\ref{ssec:gw} later.

The corrections $\delta \Omega$ and $\delta v$ are estimated from the
fitting parameters following Ref.~\cite{buonanno_kmpt2011}. To simplify
the discussion, let us assume that the coordinate separation evolves
according to
\begin{equation}
 \dot{d} (t) = A'_0 + A'_1 t + B' \cos ( \omega t + \phi_0 ) ,
\end{equation}
where $A'_0 + A'_1 t$ and $B' \cos ( \omega t + \phi_0 )$ are the
radiation-driven secular terms and eccentricity-driven modulation term,
respectively. Hereafter, we define the initial separation as $d \equiv d
(t=0)$. The sum of the eccentricity-driven initial radial velocity of
two neutron stars is estimated to be $B' \cos \phi_0$. Similarly, the
eccentricity contribution to radial acceleration of the binary is $- B'
\omega \sin \phi_0$, and this should amount to $- 2 \Omega \delta \Omega
d$ as understood by perturbing the orbital angular velocity in the
Newtonian equation of motion, $\ddot{d} = \Omega^2 d - m_0 / d$. These
are translated into the fitting parameters of $\dot{\Omega} (t)$ using
the fact that $\dot{d} (t) \approx e \omega d \cos ( \omega t + \phi_0
)$ and $\dot{\Omega} (t) \approx - 2 e \omega \Omega \cos ( \omega t +
\phi_0 )$ for a Newtonian orbit with $e \ll 1$. These two relations are
combined to suggest that $B' = - Bd/(2 \Omega)$ , and we finally find
\begin{align}
 \delta \Omega & = - \frac{B \omega \sin \phi_0}{4 \Omega^2} , \\
 \delta v & = \frac{B d \cos \phi_0}{2 \times 2 \Omega} ,
\end{align}
where $\delta v$ obtained in this manner is the value assigned to each
binary member rather than to the binary separation. This fact is
expressed by the first ``2'' in the denominator. Thus, both these
corrections can be applied directly to $\Omega$ and $v$ in
Sec.~\ref{sec:initial}.

\section{Demonstration} \label{sec:result}

We test the eccentricity reduction method described in this paper with
two equal-mass binary neutron stars. One is a $1.35 M_\odot$--$1.35
M_\odot$ binary with the H4 equation of state \cite{lackey_no2006}
denoted as the H4-135 family, and the other is a $1.4 M_\odot$--$1.4
M_\odot$ binary with the APR4 equation of state \cite{akmal_pr1998}
denoted as the APR4-14 family. These equations of state are modeled by
piecewise polytropes that approximate nuclear-theory-based ones
accurately in an analytic manner \cite{read_lof2009}. The radius of a
$1.35 M_\odot$ neutron star with H4 is 13.6 km and that of $1.4 M_\odot$
with APR4 is 11.1 km. The maximum masses of a cold, spherical neutron
star are $2.03 M_\odot$ and $2.20 M_\odot$ for H4 and APR4,
respectively. We expect that our choices of equations of state are
irrelevant to the eccentricity reduction procedure, and plan to apply
this method to models with other equations of state such as tabulated
ones in the near future. We add an ideal-gas-like thermal correction to
the equation of state during dynamical simulations with
$\Gamma_\mathrm{th} = 1.8$ in the terminology of
Ref.~\cite{hotokezaka_kkosst2013}.

\subsection{Initial data property} \label{ssec:initprop}

\begin{table*}
 \caption{Key quantities of binary-neutron-star models. Names of models
 represent the neutron-star equation of state, gravitational mass of a
 single neutron star in isolation, and stage of eccentricity
 reduction. Specifically, QC stands for quasicircular, and IterX stands
 for the Xth iteration. The total masses of the binary at infinite
 separation, $m_0$, are $2.7 M_\odot$ and $2.8 M_\odot$ for the H4-135
 and APR4-14 families, respectively. The normalized orbital angular
 velocity $m_0 \Omega$ and approaching velocity $v$ characterize the
 initial data. The Arnowitt--Deser--Misner mass and total angular
 momentum of the system are given by $M_0$ and $J_0$, respectively, and
 the former is shown as the binding energy defined by $| M_0 - m_0
 |$. The eccentricity $e$ is estimated during the fitting procedure
 using Eq.~\eqref{eq:ecc}. Each family has common values of the orbital
 separation, $d$, as well as the size of the simulation grid, $L$, and
 finest grid resolution, $\Delta x_\mathrm{min}$, in dynamical
 simulations. The initial orbital period, gravitational-wave frequency,
 and wavelength are $\approx$ 4.4 ms, 450 Hz, and 660 km for H4-135,
 respectively, and $\approx$ 4.6 ms, 440 Hz, and 680 km for APR4-14,
 respectively.}
 \begin{tabular}{cccccc} \hline
  Model & $m_0 \Omega$ & $v$ & $| M_0 - m_0 | [M_\odot]$ & $J_0
  [M_\odot^2]$ & $e$ \\
  \hline
  \multicolumn{6}{c}{H4, $1.35 M_\odot$--$1.35 M_\odot$, $d \approx$
  52.4 km, $L \approx$ 3400 km, $\Delta x_\mathrm{min} \approx$ 330
  m} \\
  \hline
  H4-135-QC & 0.0190000 & 0 & 0.0222 & 7.661 & 0.01 \\
  H4-135-Iter1 & 0.0190373 & $-0.001044$ & 0.0220 & 7.679 & 0.004 \\ 
  H4-135-Iter2 & 0.0190616 & $-0.001039$ & 0.0220 & 7.690 & 0.002 \\ 
  H4-135-Iter3 & 0.0190658 & $-0.000848$ & 0.0220 & 7.692 & 0.0008 \\ 
  \hline
  \multicolumn{6}{c}{APR4, $1.4 M_\odot$--$1.4 M_\odot$, $d \approx$
  54.3 km, $L \approx$ 2700 km, $\Delta x_\mathrm{min} \approx$ 250
  m} \\
  \hline
  APR4-14-QC & 0.0190000 & 0 & 0.0234 & 8.225 & 0.01 \\
  APR4-14-Iter1 & 0.0190437 & $-0.001065$ & 0.0233 & 8.246 & 0.005 \\
  APR4-14-Iter2 & 0.0190749 & $-0.001068$ & 0.0232 & 8.260 & 0.002 \\
  APR4-14-Iter3 & 0.0190816 & $-0.000893$ & 0.0231 & 8.264 & 0.0008 \\
  \hline
 \end{tabular}
 \label{table:model}
\end{table*}

Key quantities of initial data are summarized in Table
\ref{table:model}. We first compute quasicircular initial data labeled
by ``QC'' tuning the orbital separation to achieve the value $m_0 \Omega
= 0.019$ of the angular velocity, written in the dimensionless form $m_0
\Omega$. In subsequent computations of low-eccentricity initial data
labeled by ``IterX,'' where X is the serial number of the iteration, the
values of $m_0 \Omega$ and $v$ are prespecified. Hence, the values of
$m_0 \Omega$ and $v$ for IterX models are exact (truncated for the
presentation) rather than ones written to significant digits.

\begin{figure}
 \includegraphics[width=90mm,clip]{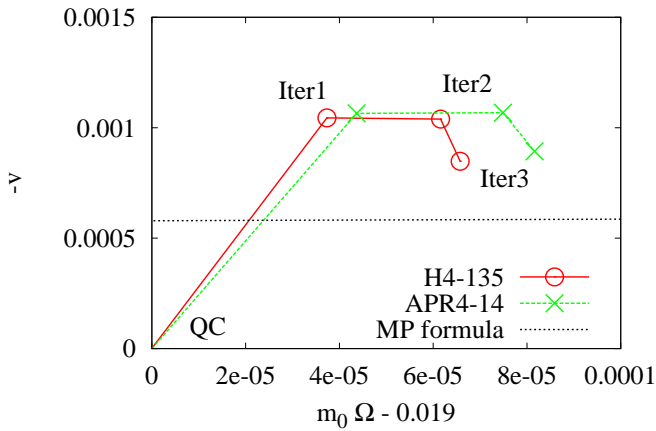} \caption{The normalized
 orbital angular velocity $m_0 \Omega$ and approaching velocity $v$ of
 initial data. The circles and crosses show the values of the H4-135 and
 APR4-14 families, respectively. We plot $m_0 \Omega$ as a difference
 from the quasicircular value 0.019 and $v$ as its negative, $-v>0$. A
 relation derived by fitting formulas proposed in
 Ref.~\cite{mroue_pfeiffer2012} is also plotted as the line labeled MP
 formula. This relation is indistinguishable from the relation derived
 by the quadrupole formula, $v = (8/5)(m_0 \Omega)^2$ for an equal-mass
 binary, if it is overplotted.} \label{fig:correction}
\end{figure}

Figure \ref{fig:correction} shows the values of $m_0 \Omega$ and $v$ for
all initial data computed in this study. As the eccentricity of Iter3
turns out to be smaller by an order of magnitude than that of QC (see
Table \ref{table:model}), this figure indicates that the eccentricity
reduction of quasicircular initial data with $m_0 \Omega = 0.019$
requires an increase of $\sim$ 0.3\%--0.5\% in the orbital angular
velocity and the incorporation of an approaching velocity with $\sim$
0.05\%--0.1\% of the speed of light. We further checked that Iter3 is
closer to hypothetical Iter4 estimated from simulations of Iter3 than to
Iter2 on this plot, and thus our eccentricity reduction procedure seems
to converge toward a circularized state. A comparison between Iter3
values and a relation due to Mrou{\'e} and Pfeiffer (MP), derived by
fitting formulas for initial data of binary black holes
\cite{mroue_pfeiffer2012}, suggests that the required orbital angular
velocity and approaching velocity are larger and smaller, respectively,
for initial data of binary neutron stars than those of binary black
holes. To confirm this, we performed another simulation of the H4-135
family with $m_0 \Omega$ and $v$ predicted by the MP formulas
\cite{mroue_pfeiffer2012} with the same coordinate separation and found
that the eccentricity is $\sim$ 0.0015. Note, however, that the meaning
of coordinate separations is not rigorously the same due to different
topology of spacetimes, whereas the gauge conditions are the same. In
particular, this could lead to a large relative difference within the
small range of $m_0 \Omega$ and $v$ relevant here. Still, the
eccentricity may be expressed as a Euclidean distance on a properly
rescaled $( m_0 \Omega , v )$ plane (see Eq.~(22) of
Ref.~\cite{mroue_pfeiffer2012}). This suggests that the fitting formulas
do not produce low-eccentricity initial data with $e \lesssim$ 0.001 for
binary neutron stars with $m_0 \Omega \approx 0.019$, although they may
be used to obtain first-trial initial data that are much better than
quasicircular initial data.

Table \ref{table:model} also shows the binding energy and total angular
momentum of binary initial data. The Arnowitt--Deser--Misner mass $M_0$
and total angular momentum $J_0$ are computed by volume integrals
\cite{gourgoulhon_gtmb2001,taniguchi_gourgoulhon2002,taniguchi_gourgoulhon2003,taniguchi_shibata2010},
and the binding energy is defined by $| M_0 - m_0 |$. It appears from
the table that the energy and angular momentum of quasicircular orbits
are too small to achieve low eccentricity. Order-of-magnitude estimates
suggest that the dominant contributions to increases of $M_0$ and $J_0$
during the eccentricity reduction come from the increase of $m_0
\Omega$, whereas the contribution of $v$ to the energy is negligible.

The global quantities can be compared with post-Newtonian formulas for
binary neutron stars (see Appendix \ref{app:pn}). Here, we include
point-particle contributions up to fourth post-Newtonian order
\cite{blanchet2014} and finite-size contributions up to first
post-Newtonian order to linear quadrupolar tidal deformation
\cite{vines_flanagan2013}. The formulas predict $( | M_0 - m_0 | , J_0
)$ at $m_0 \Omega = 0.019$ to be $( 0.0223 M_\odot , 7.687 M_\odot^2 )$
and $( 0.0232 M_\odot , 8.261 M_\odot^2 )$ for H4-135 and APR4-14,
respectively. Comparing these values with those shown in Table
\ref{table:model}, it is found that the eccentricity reduction improves
agreement of the angular momentum from $\sim$ 0.5\% of QC to $\sim$
0.1\% of Iter3. Because the post-Newtonian approximation is expected to
be an excellent approximation for the distant orbit computed here, this
improvement suggests that low-eccentricity initial data are more
accurate quasiequilibrium states than quasicircular ones.

\subsection{Orbital evolution and eccentricity} \label{ssec:evol}

Before presenting the results, we briefly summarize the setup of
numerical simulations. All the simulations in each model family are
performed with a fixed size of the computational domain,
$L$. Specifically, $L$ is $\approx$ 3400 km and 2600 km for the H4-135
and APR4-14 families, respectively. Computational domains of each
simulation consist of five coarser domains, which are centered at the
center of mass of the binary, and two sets of four finer domains, which
follow the binary motion. The box size halves every refinement level, as
does the grid separation. The grid separation at the finest domain
$\Delta x_\mathrm{min}$ is $\approx$ 330 m and 250 m for H4-135 and
APR4-14 families, respectively. These relatively coarse resolutions
(see, e.g.,
Refs.~\cite{hotokezaka_kkosst2013,hotokezaka_ks2013,hotokezaka_kkmsst2013})
are not the product of compromise but are chosen to show that our
eccentricity reduction works with low computational cost. To ensure that
the eccentricity observed in simulations is independent of the grid
resolution, we also perform simulations with $\Delta x_\mathrm{min}
\approx$ 270 m (20\% finer in terms of $1/\Delta x$) and 220 m (50\%
finer) for H4-135-QC and H4-135-Iter3.

\begin{figure*}
 \begin{tabular}{cc}
  \includegraphics[width=90mm,clip]{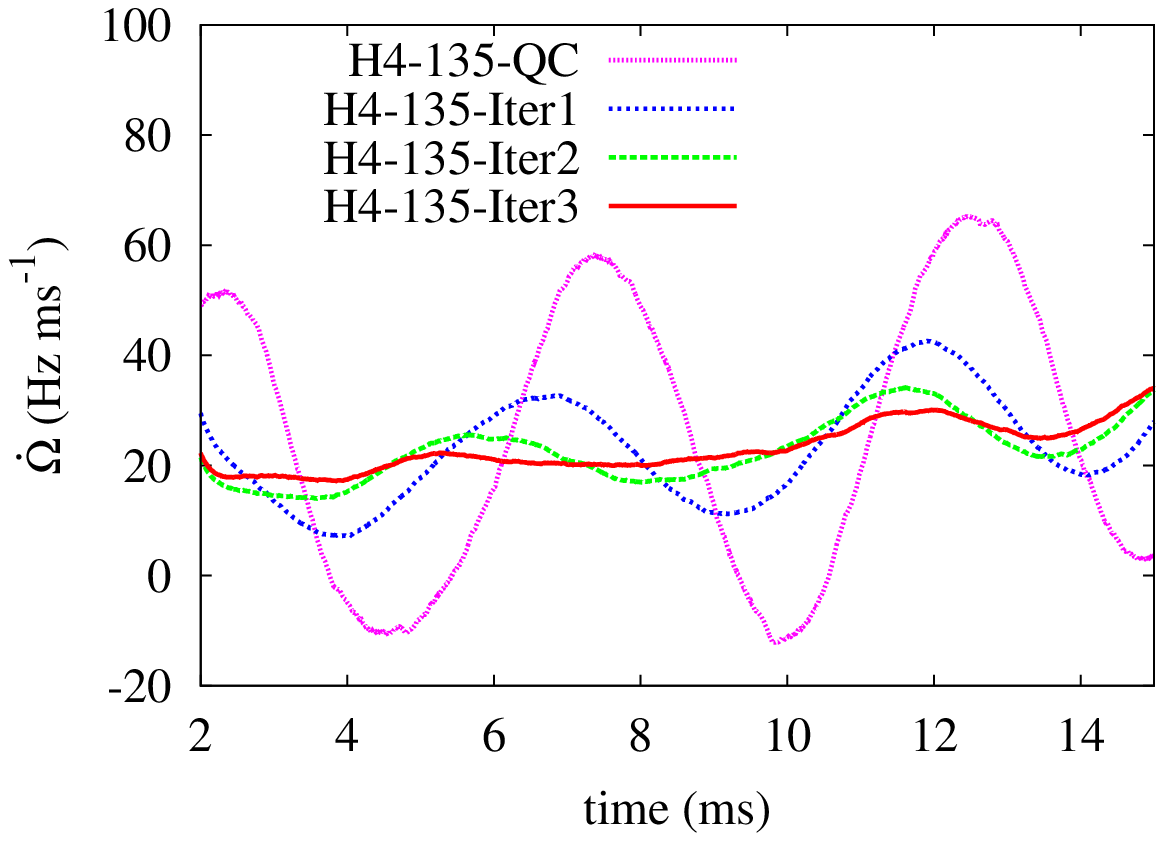} &
  \includegraphics[width=90mm,clip]{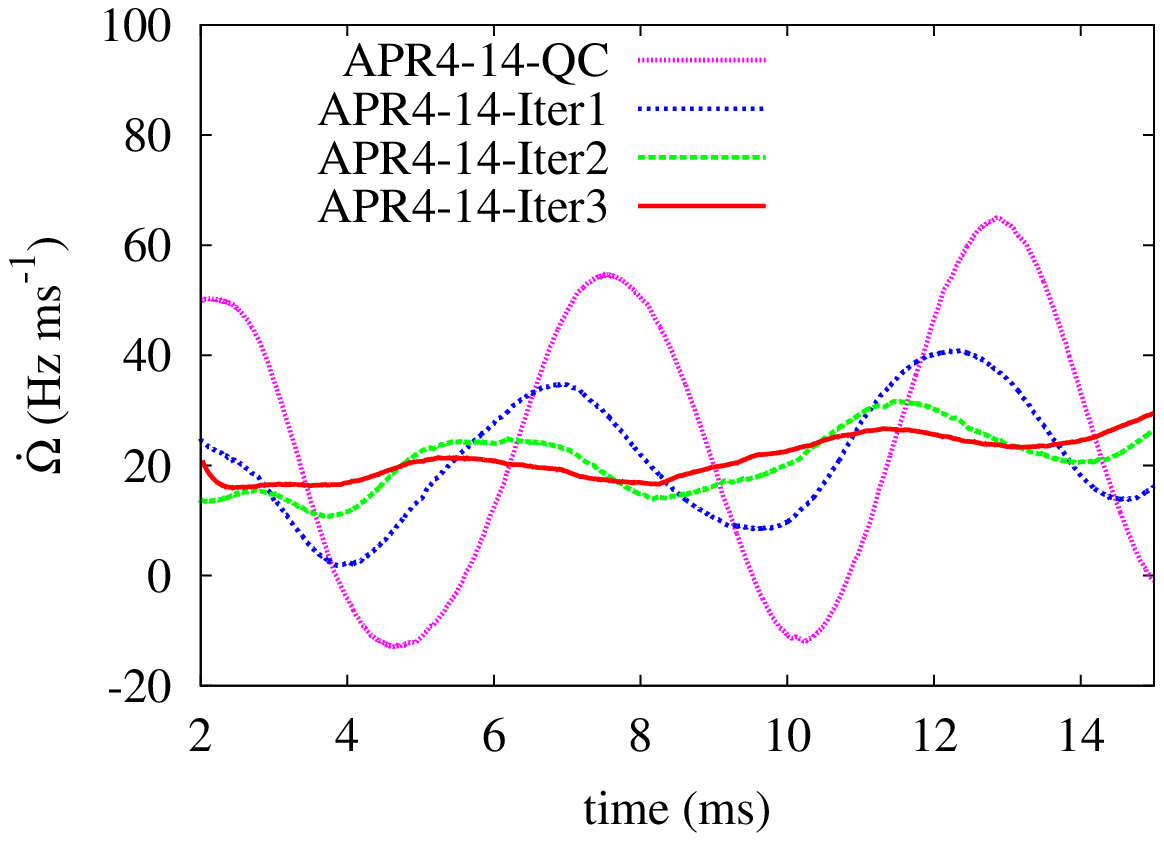}
 \end{tabular}
 \caption{The time derivative of coordinate orbital angular velocity,
 $\dot{\Omega} (t)$, of the H4-135 (left) and APR4-14 (right) families
 during the initial $\sim$ 3--4 orbits. In these plots, curves after
 taking the average over 100--200 time steps are shown. We exclude
 initial 2 ms from these plots, because data are not available for some
 models due to long averaging intervals.} \label{fig:dotomega}
\end{figure*}

First of all, Fig.~\ref{fig:dotomega} shows $\dot{\Omega} (t)$ used to
estimate corrections to initial data in the initial epoch of
simulations. This figure shows that the amplitude of modulations in
$\dot{\Omega} (t)$ decreases monotonically as the iterative eccentricity
reduction proceeds. In particular, $\dot{\Omega} (t)$ of H4-135-QC and
APR-14-QC become negative around their local minima, because the
modulation amplitude is larger than the orbital average of $\dot{\Omega}
(t)$. Negative values of $\dot{\Omega} (t)$ do not occur for all the
other initial data, because our eccentricity reduction method succeeds
in reducing the modulation in $\dot{\Omega} (t)$. It would be important,
however, to check whether our method also diminishes eccentricity-driven
modulations in other quantities that are not directly involved in the
eccentricity reduction procedure.

\begin{figure*}
 \begin{tabular}{cc}
  \includegraphics[width=90mm,clip]{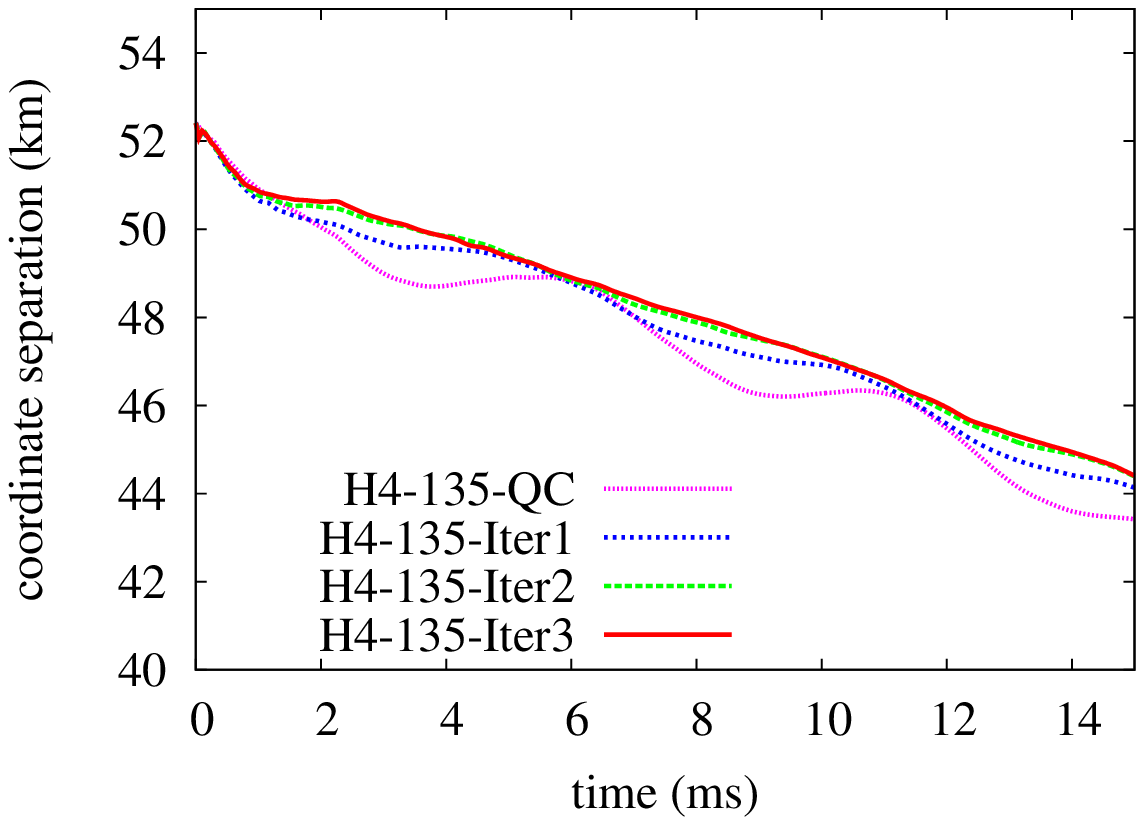} &
  \includegraphics[width=90mm,clip]{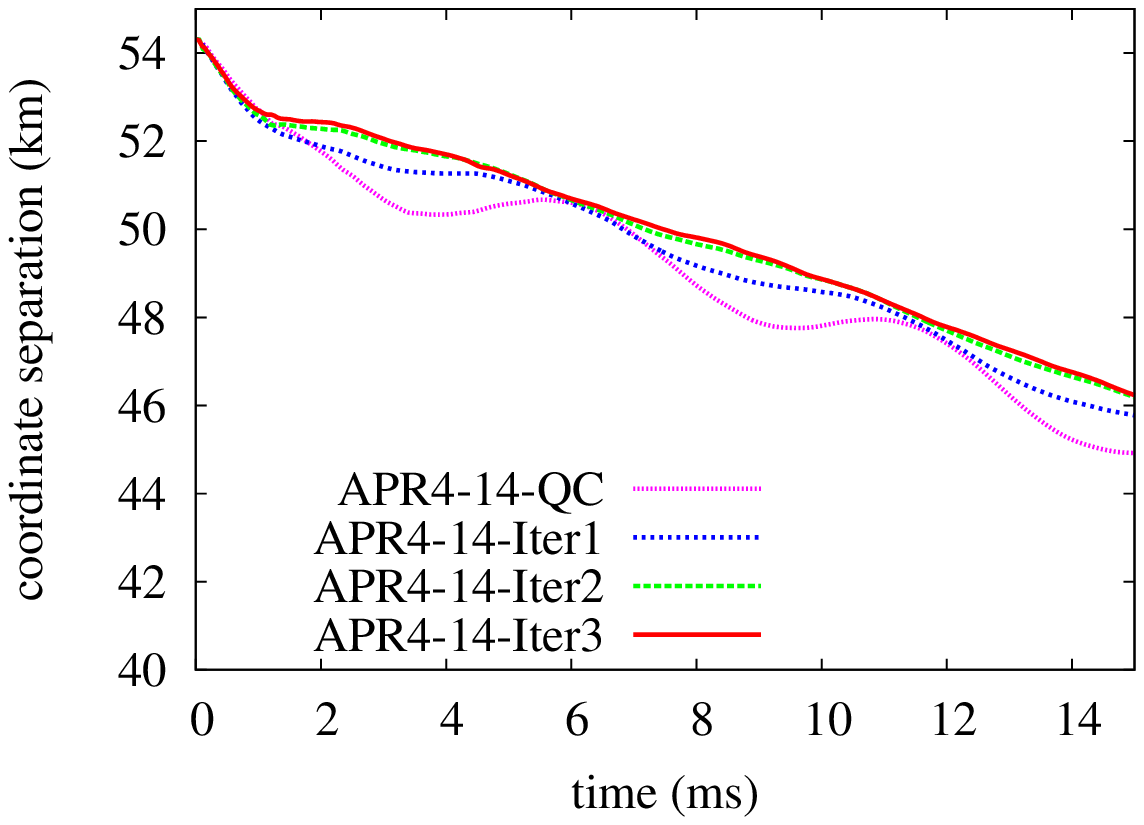}
 \end{tabular}
 \caption{The orbital evolution of the H4-135 (left) and APR4-14 (right)
 families during the initial $\sim$ 3--4 orbits. In these plots, Bezier
 curves are shown instead of raw data to focus on the
 eccentricity-driven modulation by discarding high-frequency
 oscillations associated with the finite grid separation (see
 Sec.~\ref{ssec:findorb}). We note that the Bezier smoothing is
 performed only for the presentation and is never adopted during the
 eccentricity reduction. The data averaged over 100--200 time steps
 behave very similarly to the Bezier curves, but we do not plot the
 former here, because some portion of initial abrupt decrease at
 $\lesssim$ 1 ms is masked by the averaging time interval.}
 \label{fig:orbini}
\end{figure*}

Figure \ref{fig:orbini} compares the orbital evolution of all the models
in the initial epoch of simulations. Modulations with amplitude $\sim$
0.5--1 km are observed in both H4-135-QC and APR4-14-QC, and the
coordinate separations do not even decrease monotonically. These two
models clearly exhibit the eccentricity inherent in quasicircular
initial data. The modulations decay as the iterative eccentricity
reduction proceeds, and it would not be easy to find oscillations of
either H4-135-Iter3 or APR4-14-Iter3 in these plots by eye. This sharp
decline in modulation is clear evidence of the efficacy of our
eccentricity reduction method.

The eccentricity of initial data is evaluated by Eq.~\eqref{eq:ecc} and
shown in Table \ref{table:model}. These values indicate that each
iterative correction reduces the eccentricity by a factor of 2--3. As a
result, three successive iterations reduce the eccentricity, $e \sim$
0.01, of QC to $e \lesssim$ 0.001. Because this reduction factor is
common to both the H4-135 and APR4-14 families, we believe it will not
depend on the compactness of the neutron star. It could, however, vary
with the mass ratio.

\begin{figure}
 \includegraphics[width=90mm,clip]{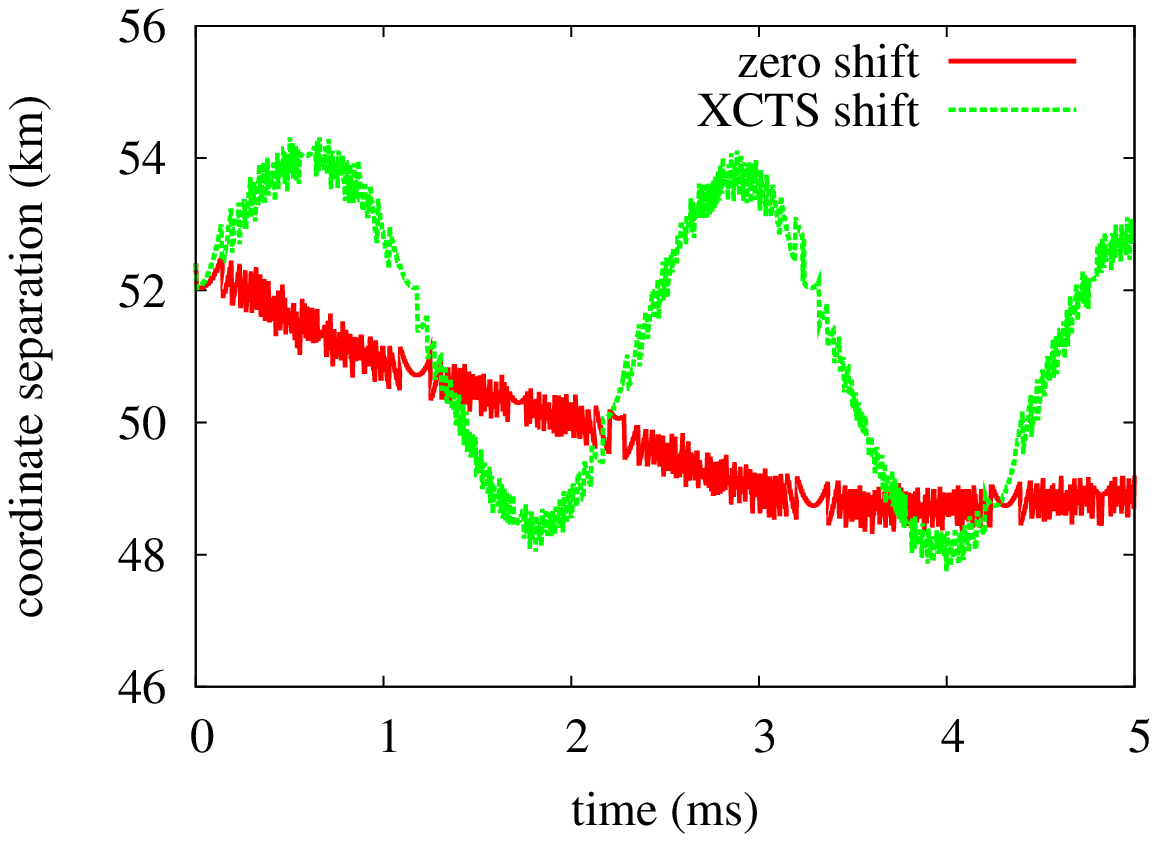} \caption{The orbital
 evolution of H4-135-QC during an initial orbit. The solid curve (zero
 shift) is the same as shown in Fig.~\ref{fig:orbini}, and derived
 giving initial data of gauge variables by $\alpha = W$, $\beta^i = 0$,
 and $B^i = 0$. The dashed oscillating curve (XCTS shift, where XCTS
 stands for extended conformal thin-sandwich) is derived giving $\alpha$
 and $\beta^i$ obtained in the extended conformal thin-sandwich
 formulation, and $B^i = 0$. In this figure, raw data are plotted to
 confirm that the initial trend is not an artifact of Bezier
 curves. This also indicates the uncertainty in determining the location
 of neutron stars by the maximum-density point, $x_\mathrm{NS}^i$.}
 \label{fig:gauge}
\end{figure}

Aside from the eccentricity-driven modulation, an abrupt decrease of the
coordinate orbital separation is found for all the models during the
initial $\sim$ 1 ms, which is approximately one-fifth of the initial
orbital period. Similar behavior is also observed in simulations of
(spinning) binary black holes based on the moving puncture method
\cite{tichy_marronetti2011}. We speculate that this decrease is due at
least partly to the initial choice of the shift vector, $\beta^i = 0$,
while the relaxation from the spatial conformal flatness may also play a
role. Figure \ref{fig:gauge} shows the orbital evolution during an
initial orbit obtained with initial shift vectors $\beta^i = 0$ and
$\beta^i$ obtained in the extended conformal thin-sandwich
formulation. The abrupt decrease is not observed with nonzero $\beta^i$,
and instead the coordinate orbital separation increases due to another
oscillation with approximately twice the frequency of the orbital
one. Excitation of second harmonics in the coordinate orbital evolution
due to $\beta^i$ obtained in the extended conformal thin-sandwich
formulation has been found in simulations of binary black holes (see
Appendix B of Ref.~\cite{yamamoto_st2008}). Because gauge-invariant
quantities such as gravitational waves depend only weakly on the choice
of initial shift vector, the harmonics should entirely be gauge
artifacts. We also find that the coordinate evolution including the
initial abrupt decrease is approximately unchanged when the harmonic
gauge condition is adopted with the shift vector initialized by $\beta^i
= 0$. Taking the different orbital evolution after this initial
transient among the models shown in Fig.~\ref{fig:orbini} into account,
we speculate that the abrupt decrease is caused by the relaxation of the
shift vector from $\beta^i = 0$, and not specific to the moving puncture
gauge. If this abrupt decrease is eliminated by a sophisticated gauge
choice, we would be able to use the numerical evolution data near $t=0$
for the eccentricity reduction, and this may improve the efficiency. By
contrast, the eccentricity reduction becomes difficult due to the second
harmonics if we adopt $\beta^i$ obtained in the extended conformal
thin-sandwich formulation as initial values.

The periastron advance of binary neutron stars does not deviate
significantly from that of binary black holes. The ratio of the radial
frequency $\omega$ and orbital frequency $\Omega$ can be estimated in
the fitting, and the initial value of $\Omega$ is typically larger than
$\omega$ by $\sim$ 20\%--25\%. At the same time, the value of the
angular velocity is estimated to be larger by $\sim$ 10\% on average
than its initial value during the fitting time interval, $[0.5P:3P]$,
due to gravitational radiation reaction. Thus, the angular frequency is
estimated to be larger by $\sim$ 30\%--35\% than the radial
frequency. This fraction is consistent with results for binary black
holes \cite{mroue_pkt2010,letiec_mbbpst2011}. Expected values for a test
particle in the Schwarzschild spacetime are 32\% and 39\% for $m_0
\Omega = 0.019$ and 0.023 (value at $t \approx 3P$), respectively, and a
third-order post-Newtonian calculation for an equal-mass binary gives
29\% and 34\% for $m_0 \Omega = 0.019$ and 0.023, respectively. Results
of binary black hole simulations lie between these predictions, and
binary neutron stars do not show significant deviations.\footnote{The
periastron advance is also studied in \cite{foucart_bdgkmmpss2013} for a
black hole-neutron star binary, but their binary configuration is chosen
so that finite-size effects do not play a role during the
evolution. Their results do not exhibit finite-size effects on the
periastron advance as expected.} More accurate estimation of a possible
finite-size effect on the periastron advance will be an interesting
topic for the future study.

\begin{figure*}
 \begin{tabular}{cc}
  \includegraphics[width=90mm,clip]{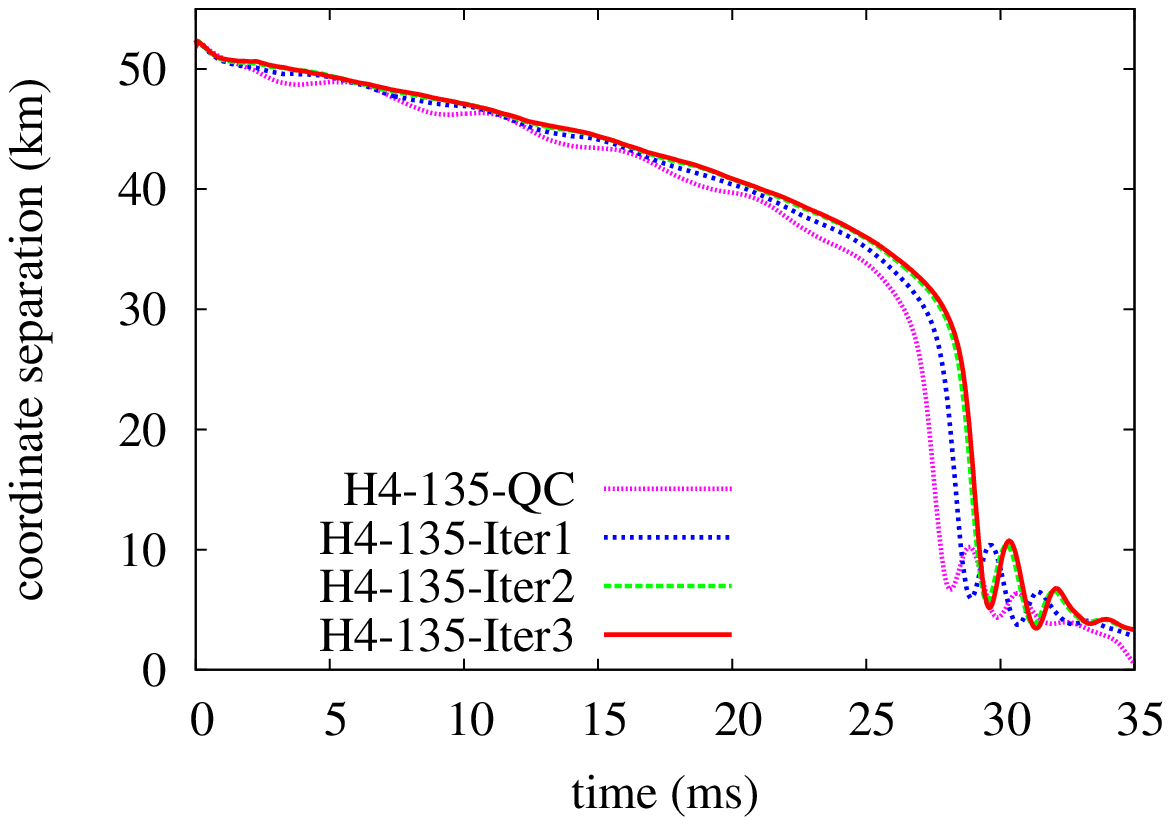} &
  \includegraphics[width=90mm,clip]{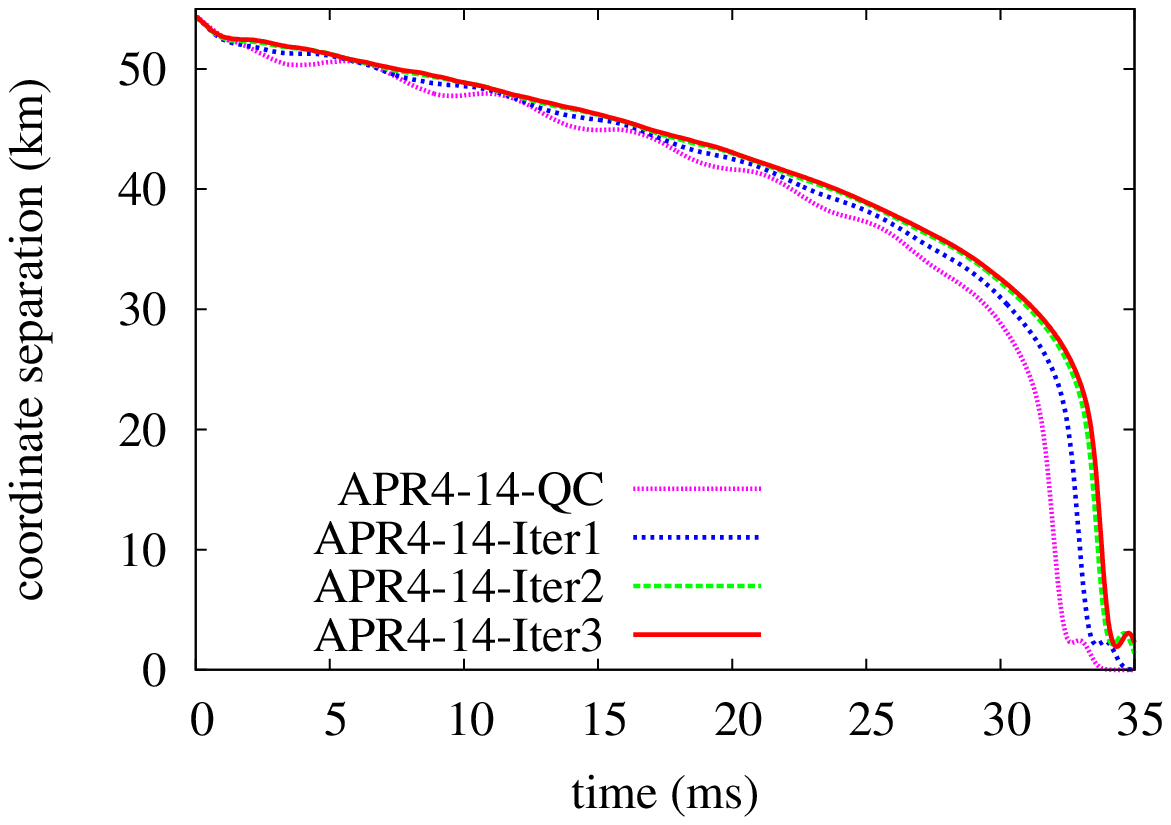}
 \end{tabular}
 \caption{The same as Fig.~\ref{fig:orbini} but up to the
 merger. Oscillations after $\sim$ 30 ms show dynamics of remnant
 massive neutron stars with double-core structures
 \cite{hotokezaka_kkmsst2013} and thus are irrelevant to the orbital
 motion.} \label{fig:orbmer}
\end{figure*}

Figure \ref{fig:orbmer} shows the orbital evolution up to the merger for
all the models. It is observed that the eccentricity-driven modulation
is very small during the entire inspiral phase for Iter3, whereas QC
shows the modulation even right before the plunge. The eccentricity of
quasicircular initial data should decrease as a result of gravitational
radiation, but the rate is only $e (t) \propto [ d (t) ]^{19/12}$ at $e
\ll 1$ in the quadrupole approximation. Hence, the expected decrease is
only by a factor of $\sim$ 4 for the evolution (say) from $15 m_0$ to $6
m_0$, and quasicircular initial data with $\sim 0.01$ will never achieve
a low-eccentricity inspiral with $e \lesssim 0.001$. Both analytic
estimation and numerical results indicate that the eccentricity
reduction has to be performed by improving initial data if we want to
obtain the low-eccentricity, $e \lesssim 0.001$ inspiral with current
computational resources.

The time to merger increases as the eccentricity decreases (see
Fig.~\ref{fig:orbmer}). While it is difficult to define the merger time
in a physical manner from the orbital evolution, the time to merger
seems to be longer by $\sim$ 2 ms for Iter3 than for QC. This cannot be
explained by strong gravitational radiation from an eccentric binary,
because the quadrupole formula predicts that the time to merger from a
given semimajor axis is proportional to $1 - (157/43) e^2$ at leading
order of $e$ \cite{peters1964}: The eccentricity enters only through its
squared value.\footnote{The absence of the first-order term is due to
averaging over the period in the derivation of emission rates.} The
actual difference in the time to merger is, however, $\sim$ 5\% between
QC with $e \sim$ 0.01 and Iter3 with $\sim$ 0.001 and is thus too
large. Instead, this difference is ascribed to the initial semimajor
axis of the binary. Numerical results suggest that apastron distances
are approximately the same irrespective of the eccentricity reduction,
and the semimajor axis is shorter by $1 - e$. Thus, the time to merger
should be proportional to $1 - 4e$ at leading order of $e$. This
approximately explains the difference of the time to merger shown in
Fig.~\ref{fig:orbmer}. This feature is consistent with findings in
Ref.~\cite{pfeiffer_bklls2007} that modulation-free parts of the orbital
evolution and gravitational radiation are approximately independent of
small eccentricity once appropriate time shifts are applied. This
argument also suggests that quasicircular initial data computed assuming
helical symmetry represent a binary configuration at the apastron. That
is, the coordinate separation $d$ specified in the initial data
computation is the apastron distance.

\subsection{Gravitational waves} \label{ssec:gw}

The binary's dynamical behavior is more reliably observed using
gravitational waves, which are gauge invariant at least
asymptotically. In {\small SACRA}, gravitational waves are extracted via
the Weyl scalar $\Psi_4$ projected onto spin-weighted spherical
harmonics at finite coordinate radii. We focus only on the $( \ell , m )
= (2,2)$ mode in this study and denote its coefficient as
$\Psi_{4,22}$. Although we extrapolate $\Psi_4$ to null infinity when we
compute gravitational waves as transverse-traceless portions of the
metric, we do not perform the extrapolation when we present $\Psi_4$
itself and instead show results obtained by extraction at $r_\mathrm{ex}
\approx$ 300 km. We checked that $\Psi_4$ in the inspiral
phase\footnote{Postmerger gravitational waves from the remnant massive
neutron star are not extracted reliably at larger radii, because the
grid resolution is not enough at the distant region to cover one
gravitational wavelength by $\sim$ 10 points due to high frequency and
the coarse resolution adopted for the eccentricity reduction.} depends
only weakly on the extraction radius once a time shift and amplitude
scaling are performed according to the different extraction radii. We
present gravitational-wave quantities as a function of an approximate
retarded time defined using an approximate areal radius $D$ as
\begin{align}
 t_\mathrm{ret} & \equiv t - D - 2 m_0 \ln \left( \frac{D}{m_0} \right)
 , \label{eq:retard} \\
 D & \equiv r_\mathrm{ex} \left( 1 + \frac{m_0}{2 r_\mathrm{ex}}
 \right)^2 \label{eq:areal} .
\end{align}
The total mass $m_0$ in these definitions should have been replaced by
the Arnowitt--Deser--Misner mass, $M_0$, but we choose to avoid using
$M_0$, because this depends on the stage of the eccentricity reduction
as shown in Table \ref{table:model}. The difference between $m_0$ and
$M_0$ does not introduce appreciable effects as far as we are concerned
in this study.

\begin{figure*}
 \begin{tabular}{cc}
  \includegraphics[width=90mm,clip]{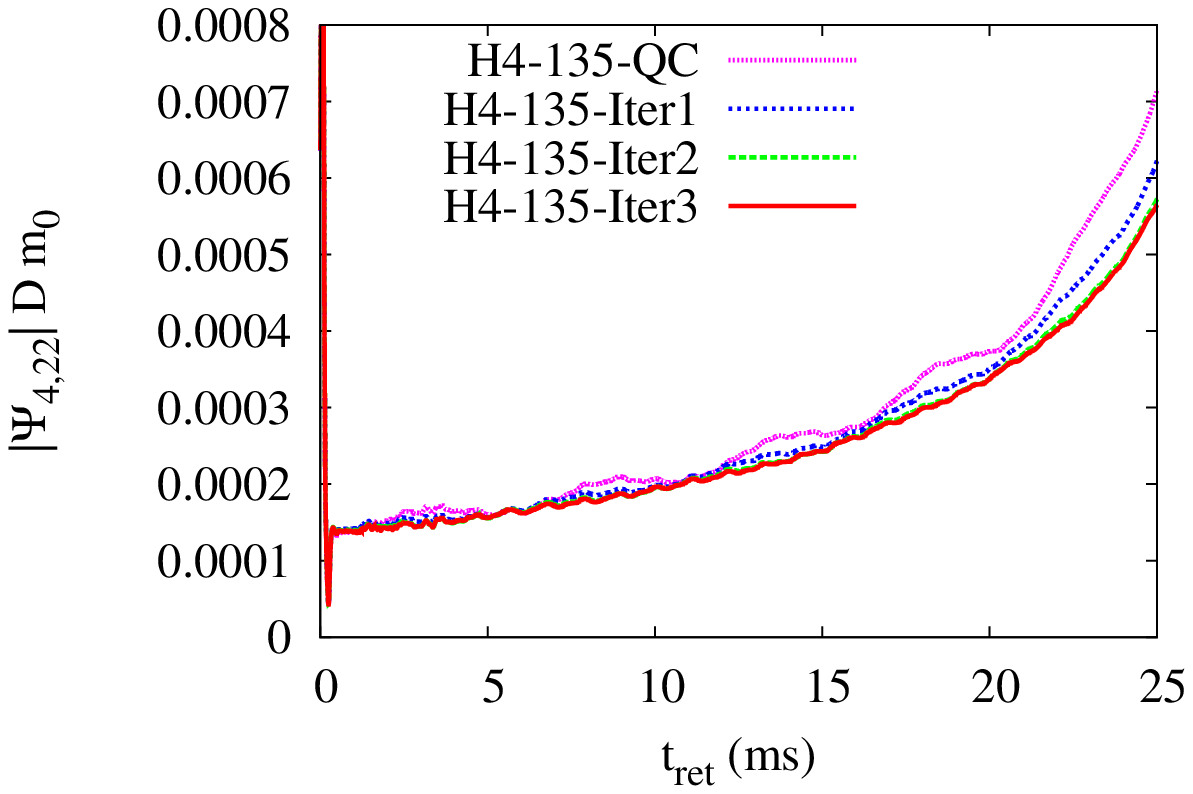} &
  \includegraphics[width=90mm,clip]{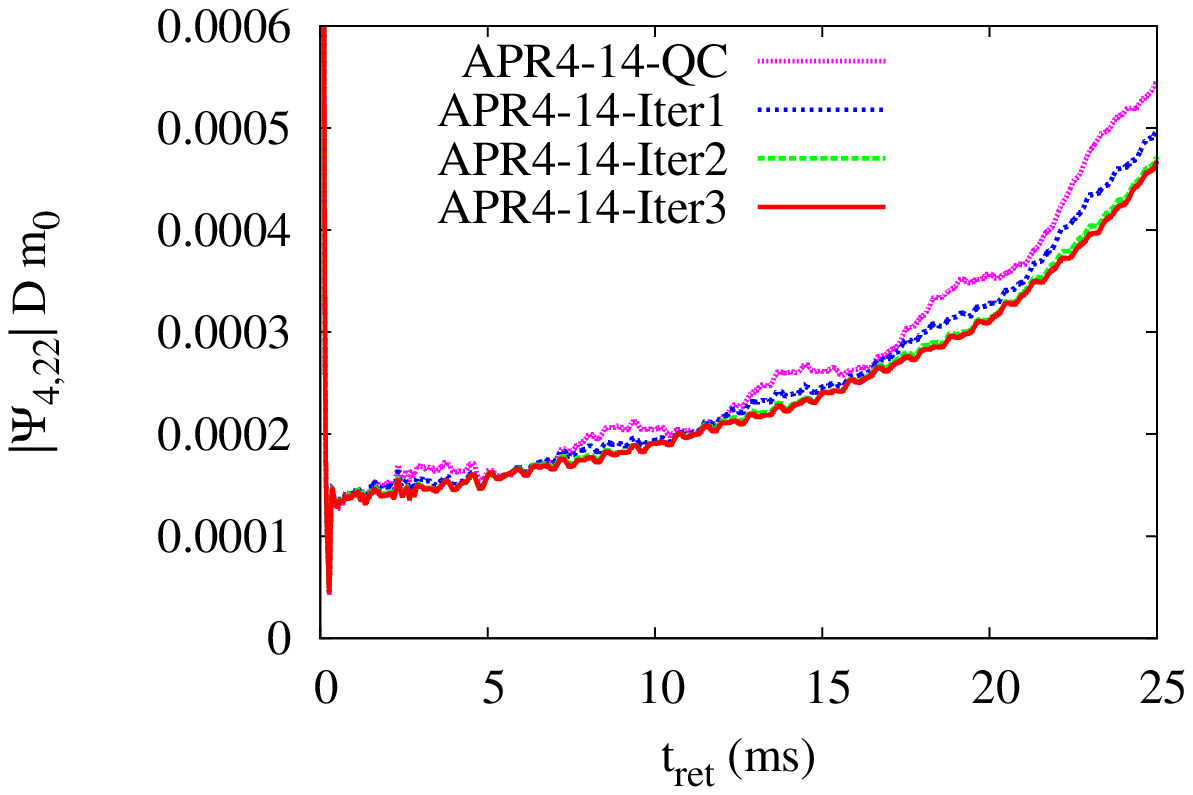}
 \end{tabular}
 \caption{The amplitude of $\Psi_{4,22}$ in the inspiral phase for the
 H4-135 (left) and APR4-14 (right) families. The amplitude is given by a
 dimensionless combination, $| \Psi_{4,22} | D m_0$. The time is given
 by an approximate retarded time, and initial $\sim$ 0.1 ms suffers from
 junk radiation in initial data.} \label{fig:psi4}
\end{figure*}

Figure \ref{fig:psi4} shows the amplitude of $\Psi_{4,22}$ in the
inspiral phase obtained from each model. Amplitudes of QC show
substantial modulations in a similar manner to the orbital evolution. By
contrast, those of IterX show only small modulations compared to QC, and
again the modulations of Iter3 are not easily detected by eye except for
small wiggles. The wiggles have higher frequency than the orbital
frequency, and they are not induced by the eccentricity. Because
gravitational-wave quantities are expected to be gauge invariant, this
figure gives a firm evidence of the success of our eccentricity
reduction. A comparison of Figs.~\ref{fig:orbmer} and \ref{fig:psi4}
suggests that the eccentricity-driven modulations are approximately in
phase for these two quantities, and thus discussions based on the
coordinate orbital separation are not entirely gauge artifacts.

\begin{figure*}
 \begin{tabular}{cc}
  \includegraphics[width=90mm,clip]{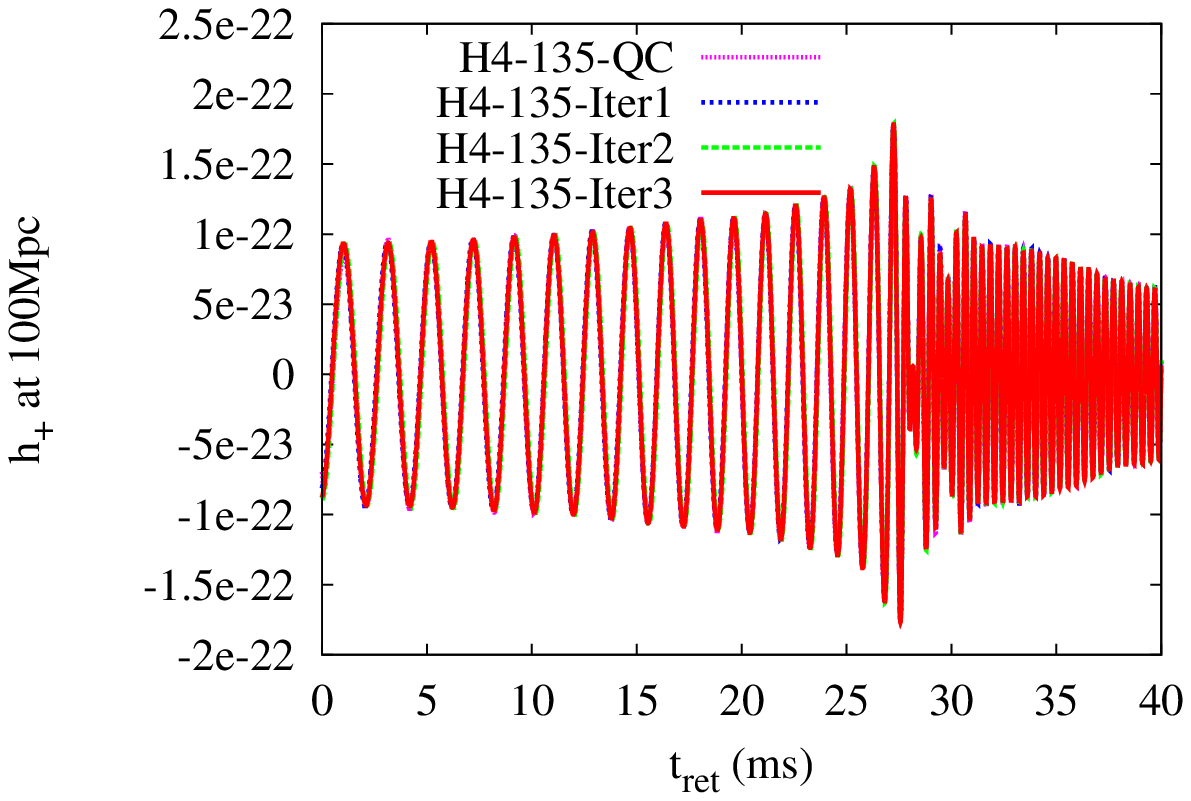} &
  \includegraphics[width=90mm,clip]{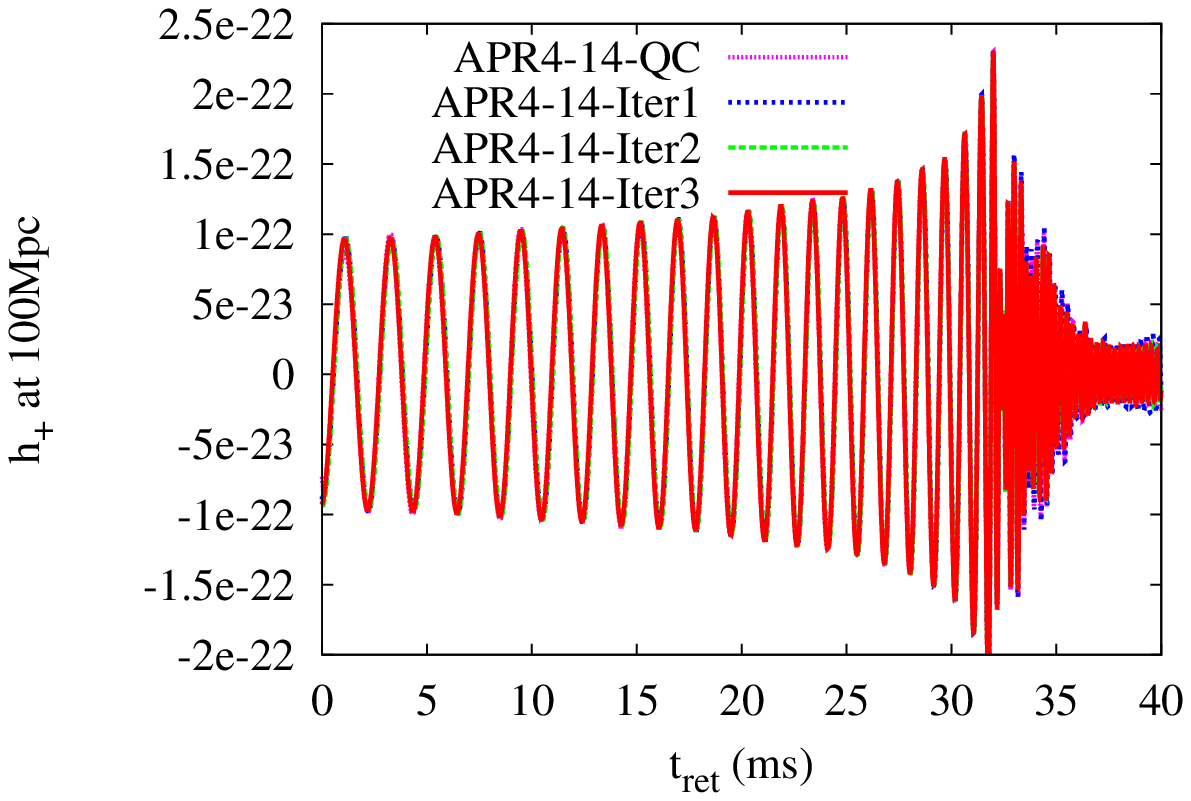}
 \end{tabular}
 \caption{Gravitational waveforms for the H4-135 (left) and APR4-14
 (right) families. The amplitude is given as that for an observer at 100
 Mpc distance along the rotational axis of the binary. The plus mode
 gravitational waves of QC are drawn as a function of the approximate
 retarded time. Those of IterX are shifted in time so that the maximum
 amplitude occurs at the same time as QC and also shifted in phase so
 that polarization states approximately agree with that of QC.}
 \label{fig:gw}
\end{figure*}

Gravitational waveforms are most important for the purpose of this
study. To extract the waveforms, we extrapolate $\Psi_4$ to null
infinity and perform a time integration as summarized in Appendix
\ref{app:gw}. We plot plus-mode gravitational waves $h_+ (t)$ in
Fig.~\ref{fig:gw} after applying time and phase shifts to align the
waveforms. The cross-mode $h_\times (t)$ shows the same behavior for the
purpose of our discussion. Here, we again note that $h_+$ and $h_\times$
are real and (negative of) imaginary parts of the $( \ell , m ) = (2,2)$
mode, respectively. This figure shows that the effect of an eccentricity
$e \lesssim 0.01$ is not easily distinguished from the shifted
gravitational waveforms, in either $\sim$ 10 orbits in the inspiral
phase or a few tens of cycles in the postmerger phase. The former is
already shown in the study of binary black holes
\cite{pfeiffer_bklls2007}, and our result confirms this for binary
neutron stars for the first time. This agreement will be made more
quantitative when we compute the overlap or mismatch of the waveforms
obtained by simulations, and we postpone such computations after
hybridization with analytic models. The visual agreement of
gravitational waveforms does not imply that quasicircular initial data
are sufficient for the construction of theoretical templates, but
properties of gravitational waves have to be investigated carefully as
we do below.

\begin{figure*}
 \begin{tabular}{cc}
  \includegraphics[width=90mm,clip]{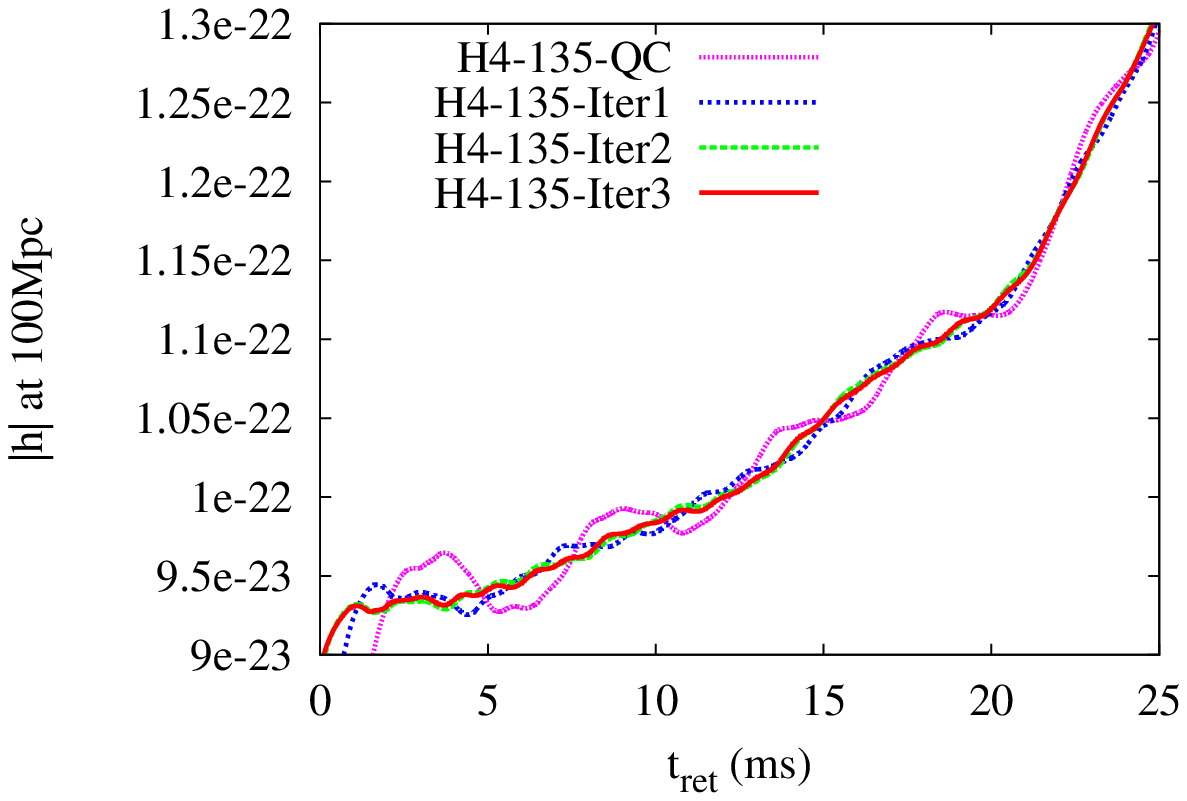} &
  \includegraphics[width=90mm,clip]{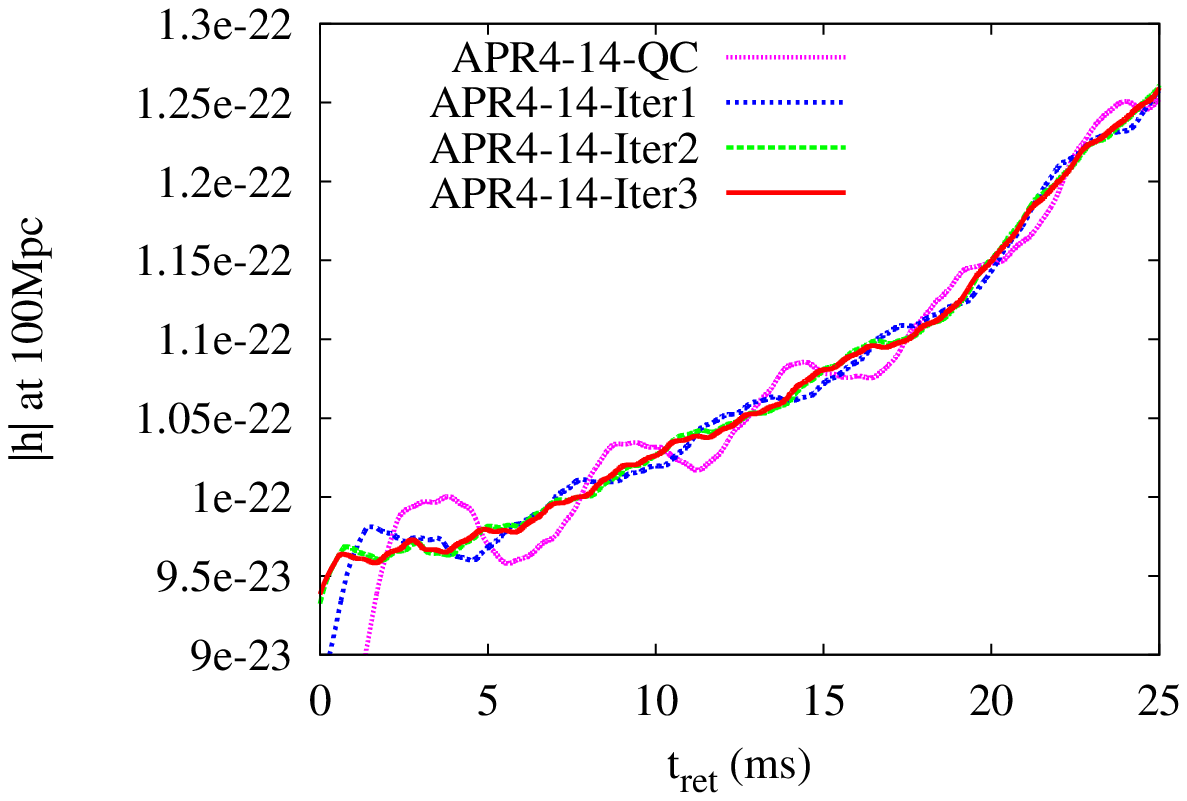}
 \end{tabular}
 \caption{Gravitational-wave amplitudes for the H4-135 (left) and
 APR4-14 (right) families in the inspiral phase. The vertical axis is
 the sum-squared amplitude measured by an observer at 100 Mpc distance
 along the rotational axis of the binary. The time axis is aligned so
 that the maximum amplitude occurs at the same time (outside these
 plots). QC and Iter1 suffer from junk radiation during the initial
 $\sim$ 1--2 ms, while junk radiation components of Iter2 and Iter3 are
 less apparent in the plots due to the time shifts.} \label{fig:amp}
\end{figure*}

We move to more in-depth discussions by decomposing the gravitational
waveforms into the amplitude and frequency. First, Fig.~\ref{fig:amp}
shows the time evolution of the amplitude defined by
\begin{equation}
 |h| (t) \equiv \sqrt{[h_+ (t)]^2 + [h_\times (t)]^2} ,
\end{equation}
and the time shift is applied in the same manner as
Fig.~\ref{fig:gw}. Although the amplitude evolution of Iter3 is very
smooth, those of QC with $e \sim$ 0.01 show the modulation with
amplitude $\sim$ 3\%--5\% throughout the inspiral phase. This is roughly
consistent with the expectation that the amplitude should vary by $(3/2)
e$ at leading order of $e$ in the quadrupole approximation. In addition,
these modulations are again in phase with the modulations observed in
the orbital phase.

\begin{figure*}
 \begin{tabular}{cc}
  \includegraphics[width=90mm,clip]{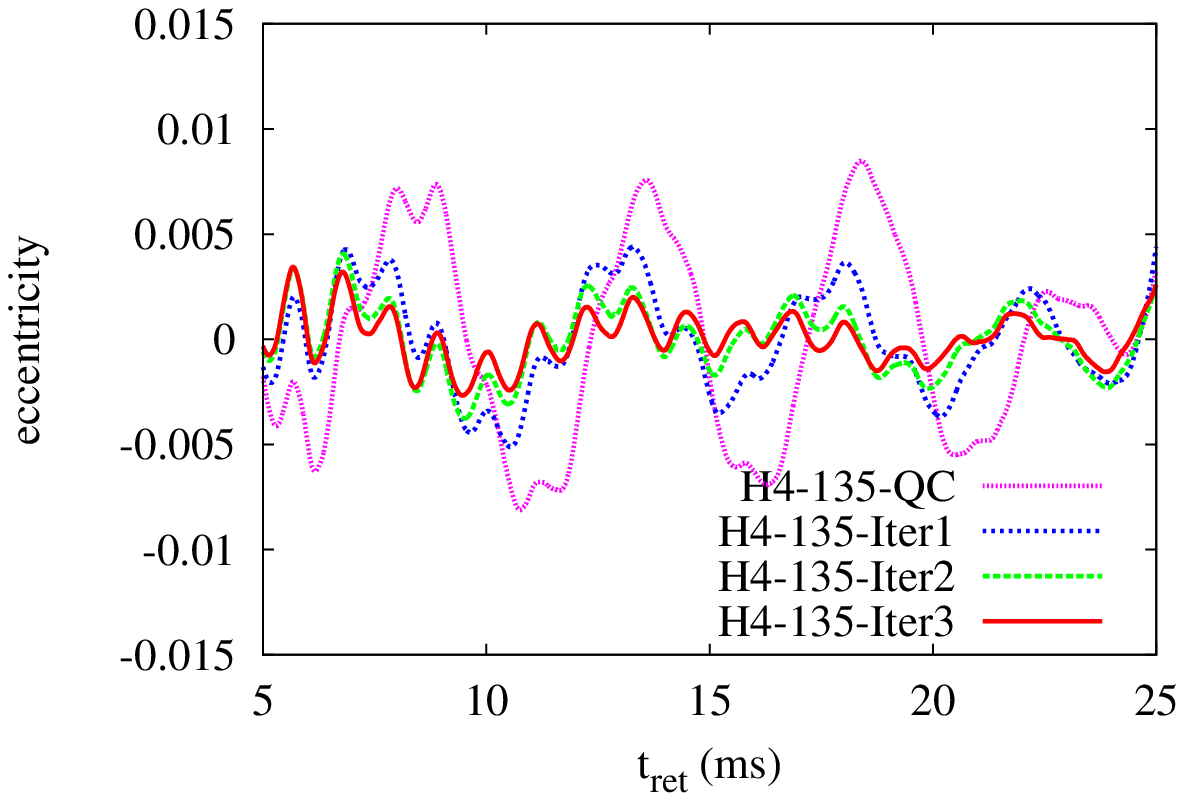} &
  \includegraphics[width=90mm,clip]{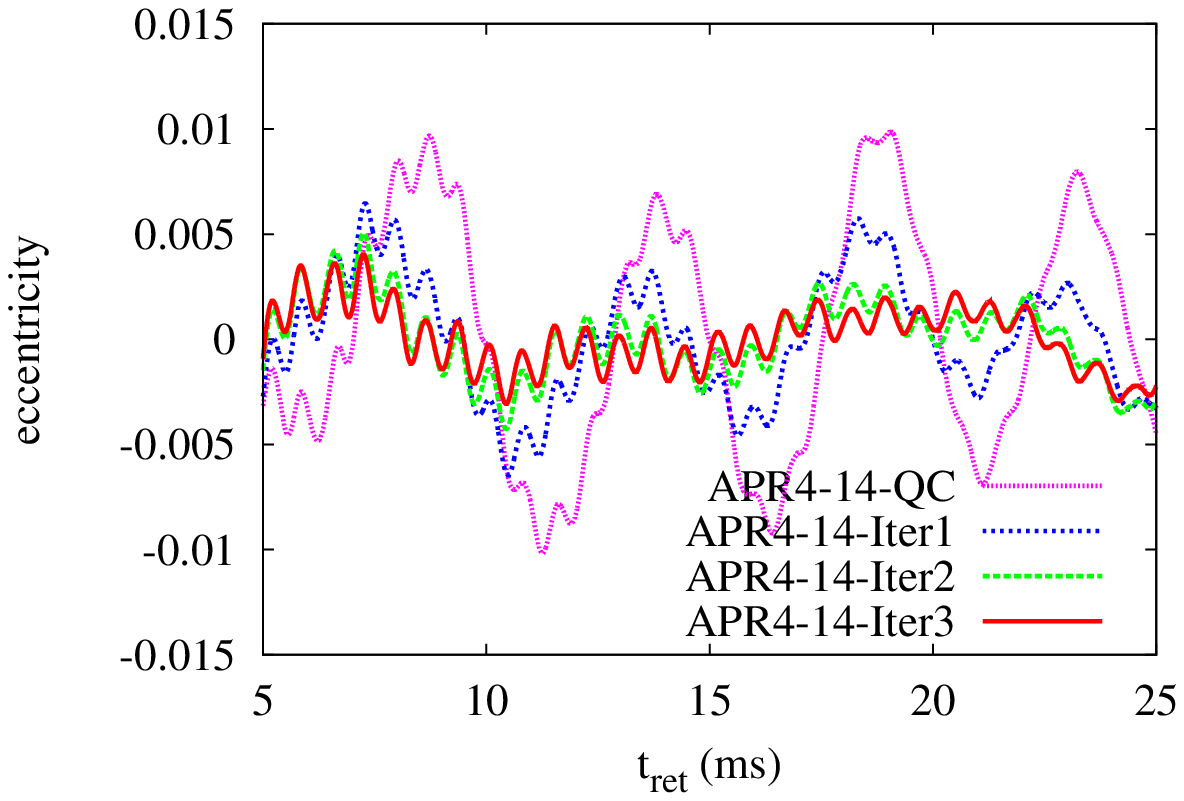}
 \end{tabular}
 \caption{Time evolution of the eccentricity estimator, $e_\mathrm{GW}
 (t)$, defined from $\Omega_\mathrm{GW} (t)$ for the H4-135 (left) and
 APR4-14 (right) families in the inspiral phase. We truncate initial
 $\sim$ 5 ms for the fitting, because junk radiation pollutes
 $\Omega_\mathrm{GW} (t)$ in the initial orbit. After $\sim$ 25 ms, the
 evolution of the binary becomes plungelike, and the eccentricity may no
 longer be defined even approximately.} \label{fig:ecc}
\end{figure*}

Next, we turn to the angular frequency of gravitational waves
$\Omega_\mathrm{GW} (t)$ and show behavior of an eccentricity estimator
instead of the angular frequency itself, which behaves very similarly to
the amplitude shown in Fig.~\ref{fig:amp}. The eccentricity estimator is
defined using an underlying smooth evolution obtained by fitting,
$\Omega_\mathrm{GW,fit} (t)$, as \cite{husa_hgsb2007}
\begin{equation}
 e_\mathrm{GW} (t) \equiv \frac{\Omega_\mathrm{GW} (t) -
  \Omega_\mathrm{GW,fit} (t)}{2 \Omega_\mathrm{GW,fit} (t)} .
\end{equation}
The fitting function is chosen to be polynomial in time,
\begin{equation}
 \Omega_\mathrm{GW,fit} (t) = a_0 + a_1 t + a_2 t^2 + a_3 t^3 + a_4 t^4
  ,
\end{equation}
and features of $e_\mathrm{GW} (t)$ depend only weakly on the specific
form of the fitting function. The fitting time interval is taken from
$\sim$ 5 ms after the initial instant to $\sim$ 1 orbit before the
merger, where the former is chosen to avoid strong unphysical effects of
junk radiation.

Figure \ref{fig:ecc} shows the time evolution of the eccentricity
estimator, $e_\mathrm{GW} (t)$. The curves of QC exhibit oscillation
components with the period $\sim$ 5 ms and amplitude $\sim$ 0.01, and
this is ascribed to the orbital eccentricity in quasicircular initial
data. This oscillation component with the period of $\sim$ 5 ms decays
as the eccentricity reduction proceeds, and the curves of Iter3 show
$\sim$ 5 ms oscillations with the amplitude of only $\sim$ 0.001. This
behavior of the eccentricity estimator is consistent with the
eccentricity evaluated in the fitting of orbital motion shown in Table
\ref{table:model}, and thus both eccentricity estimation methods may be
reliable.\footnote{See Ref.~\cite{purrer_hh2012} for conceptual
differences of these two estimators.}

When the orbital eccentricity is reduced to $\sim$ 0.001, another
oscillation component with an amplitude $\sim$ 0.001 becomes prominent
in the eccentricity estimator, $e_\mathrm{GW}$, with frequency several
times higher than the orbital and radial frequencies. Indeed, this rapid
component is not new to Iter3, and QC also exhibits this oscillatory
behavior superposed on the eccentricity-driven modulation. This
component does not converge away with increasing grid resolution as far
as we tried. We also confirmed that this oscillation cannot be ascribed
to the oscillation of neutron stars, since no oscillatory mode with the
same frequency is found up to $\ell \le 4$, including quadrupole
oscillations, in our simulations. We speculate that this oscillation
component is due to insufficient boundary conditions of {\small SACRA},
because the frequency of the oscillation changes when the location of
the outer boundary is changed,\footnote{We checked that the oscillation
is not related to the finite extraction radius,
Courant--Friedrichs--Lewy factor, strength of the Kreiss--Oliger
dissipation (as far as gravitational waves are not completely smeared
out), damping parameter $\kappa_1$, gauge parameter $\eta_s$, and
artificial atmosphere. We also checked that the oscillation does not
vanish when a simulation is performed in the BSSN formulation.} although
only slightly. We will confirm or exclude this speculation by
implementing higher-order boundary conditions \cite{hilditch_btctb2013}
in the near future. This will involve modifying the shape of the outer
boundary. Another possibility is reflection of unphysical high-frequency
radiation contained in initial data at adaptive-mesh-refinement
boundaries \cite{zlochower_pl2012}, sizes of which are proportional to
that of the outer boundary in {\small SACRA}. If this is the case,
appropriately modified gauge conditions might reduce the rapid
oscillation component \cite{etienne_bpks2014}.

\begin{figure*}
 \begin{tabular}{cc}
  \includegraphics[width=90mm,clip]{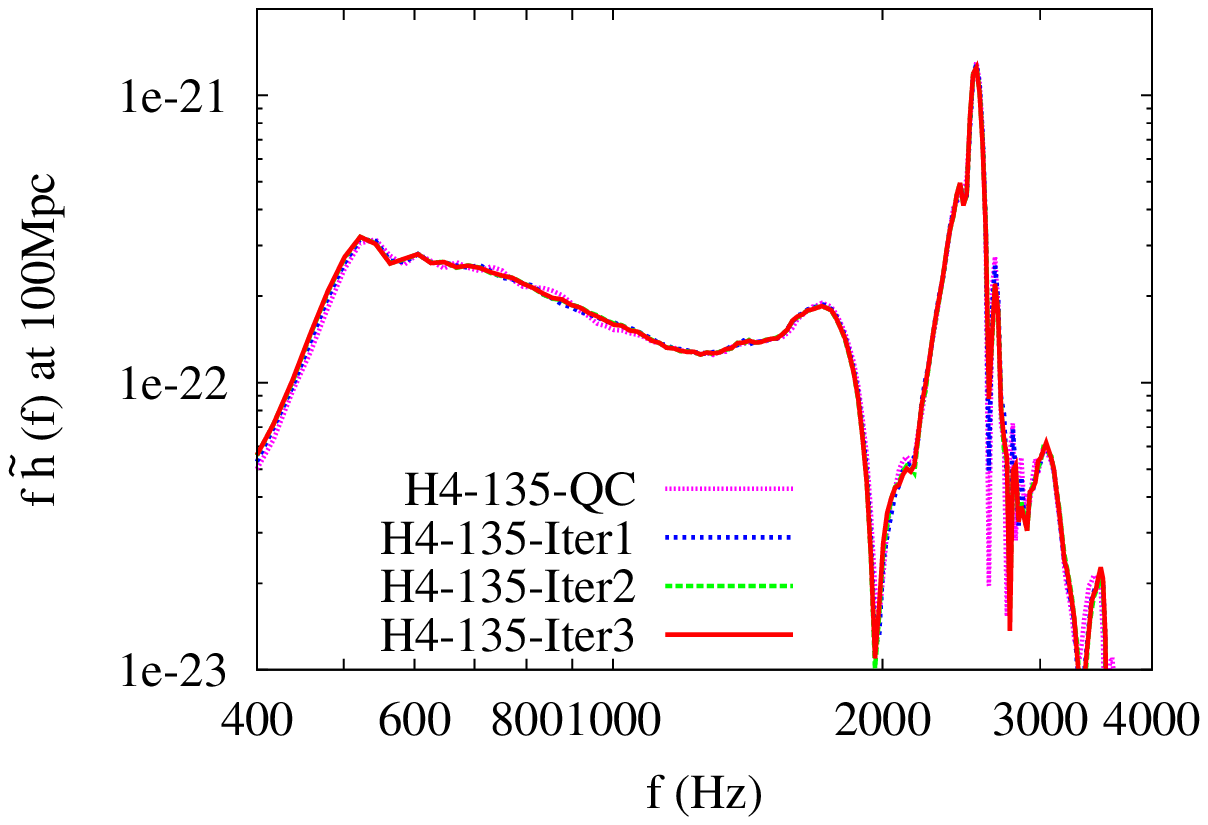} &
  \includegraphics[width=90mm,clip]{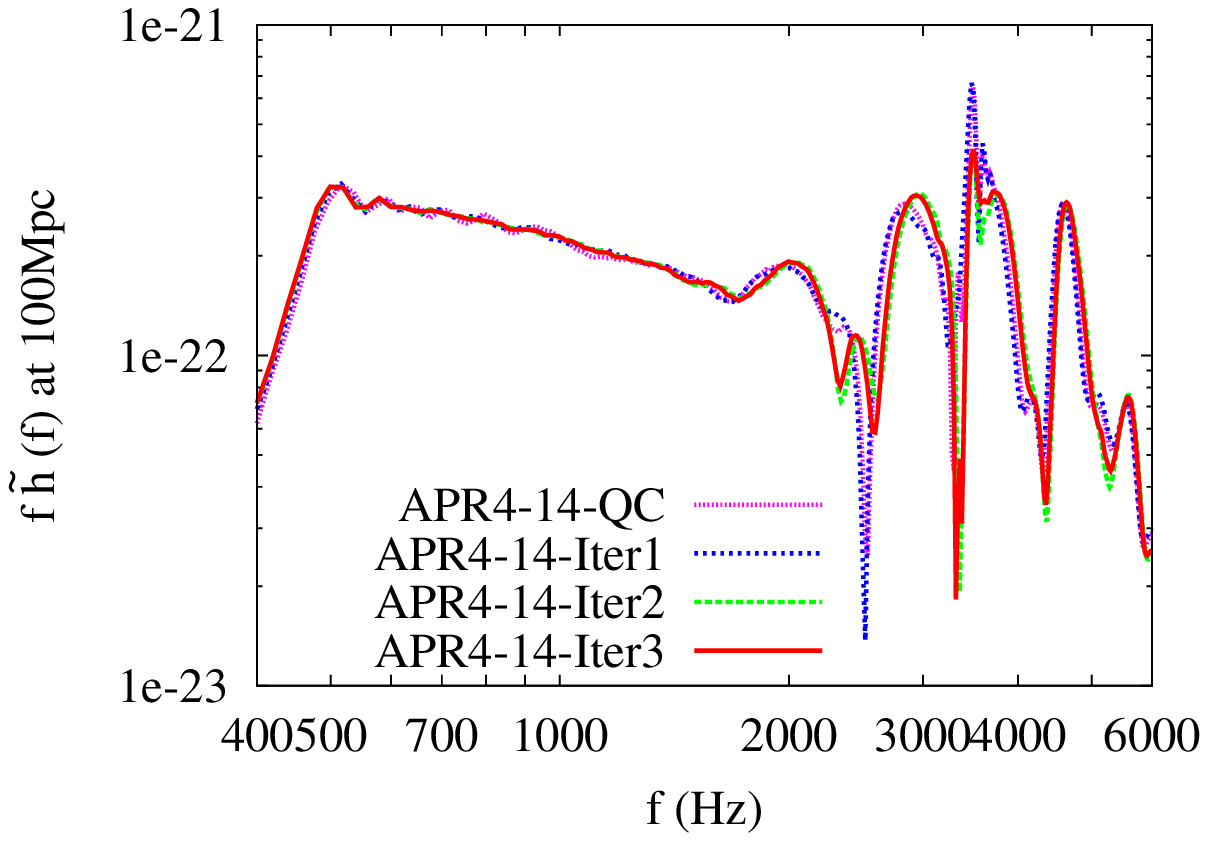}
 \end{tabular}
 \caption{Gravitational-wave spectra for the H4-135 (left) and APR4-14
 (right) families. The vertical axis is the dimensionless amplitude, $f
 \tilde{h}(f)$, measured by an observer at 100 Mpc distance along the
 rotational axis of the binary. Abrupt decreases of the amplitude below
 $\sim$ 500 Hz are simply due to the fact that our simulations begin
 there. The spectra of H4-135 below and above $\sim$ 2000 Hz are
 contributions from the inspiral and remnant massive neutron star,
 respectively, and a similar separation holds for APR4-14 but at $\sim$
 2500 Hz. Spectral peaks associated with remnant massive neutron stars
 are systematically underestimated at high-frequency sides, because we
 truncate numerical data after $\sim$ 50 ms from the initial instant (or
 $\sim$ 20 ms after the merger) when remnants are still active. The peak
 frequency, where the maximum amplitude is achieved, does not vary by
 more than $\sim$ 10 Hz with the truncation.} \label{fig:spec}
\end{figure*}

We show the spectrum of gravitational waves in
Fig.~\ref{fig:spec}. Here, the effective spectral amplitude $\tilde{h}
(f)$ is defined by
\begin{equation}
 \tilde{h} (f) \equiv \sqrt{\frac{\left[ \tilde{h}_+ (f) \right]^2 +
  \left[\tilde{h}_\times (f) \right]^2}{2}} ,
\end{equation}
where $\tilde{h}_+ (f)$ and $\tilde{h}_\times (f)$ are the Fourier
transformation of the plus and cross modes, respectively. A remnant
massive neutron star of neither H4-135-QC nor APR4-14-QC collapses to a
black hole during simulations over $\sim$ 100 ms,\footnote{The lifetimes
of remnant massive neutron stars are longer than those found in our
previous works \cite{hotokezaka_kkmsst2013} irrespective of the
eccentricity, and we speculate that accumulated constraint violation and
associated spurious dissipation of the angular momentum trigger the
early collapse in the BSSN formulation adopted there.} and therefore we
decided to stop H4-135-Iter3 and APR4-14-Iter3 only after $\sim$ 50 ms
from the initial instant (or $\sim$ 20 ms after the merger). The same
time interval of $\sim$ 50 ms is adopted among all the models for the
computation of the spectra to avoid possible biases, even if longer data
are available for some of the models. This truncation during the
lifetime of remnant massive neutron stars results in an underestimation
of the spectral peak at $\sim$ 2500 Hz and 3500 Hz for the H4-135 and
APR4-14 families, respectively, associated with postmerger activities of
the remnant massive neutron stars. This truncation mainly affects the
high-frequency side of the spectral peak, and the frequency of the
maximum amplitude does not vary by more than $\sim$ 10 Hz.

The effect of a small eccentricity, $e \lesssim 0.01$, on the spectra is
weak both for inspiral and postmerger phases, in a similar manner to the
waveforms shown in Fig.~\ref{fig:gw}. This shows that previous studies
of binary neutron stars focused on gravitational-wave spectra are not
affected significantly by the orbital eccentricity if the simulation is
sufficiently long. We find that the spectral amplitude is smoother for
Iter3 in $\sim$ 600--1500 Hz than QC, and this is likely to reflect the
reduced eccentricity. The difference is, however, relatively subtle and
easily masked by different filtering techniques in Fourier
transformation (see Appendix \ref{app:gw}). Thus, we only suggest that
low-eccentricity initial data may yield smoother spectra than
quasicircular ones. Figure \ref{fig:spec} also suggests that
characteristics of remnant massive neutron stars do not depend strongly
on the eccentricity as again expected from the postmerger agreement in
Fig.~\ref{fig:gw}.

\subsection{Convergence}

\begin{figure*}
 \begin{tabular}{ccc}
  \includegraphics[width=59mm,clip]{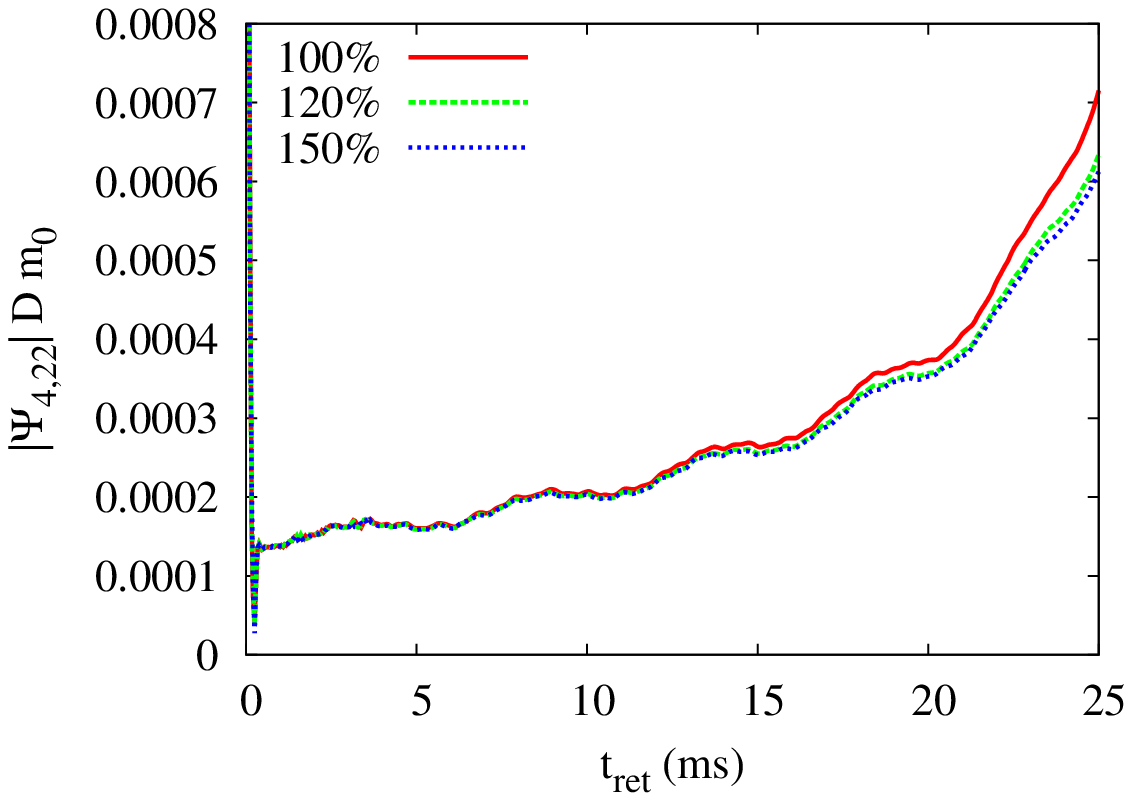} &
  \includegraphics[width=59mm,clip]{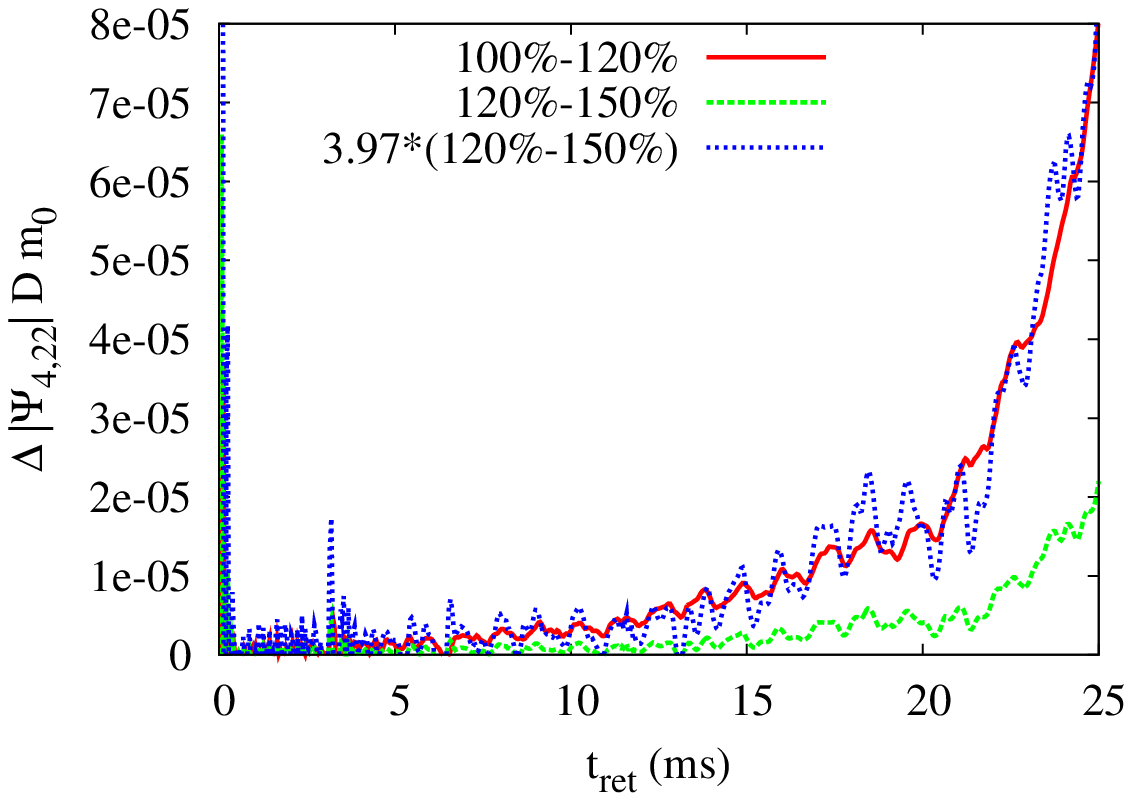} &
  \includegraphics[width=59mm,clip]{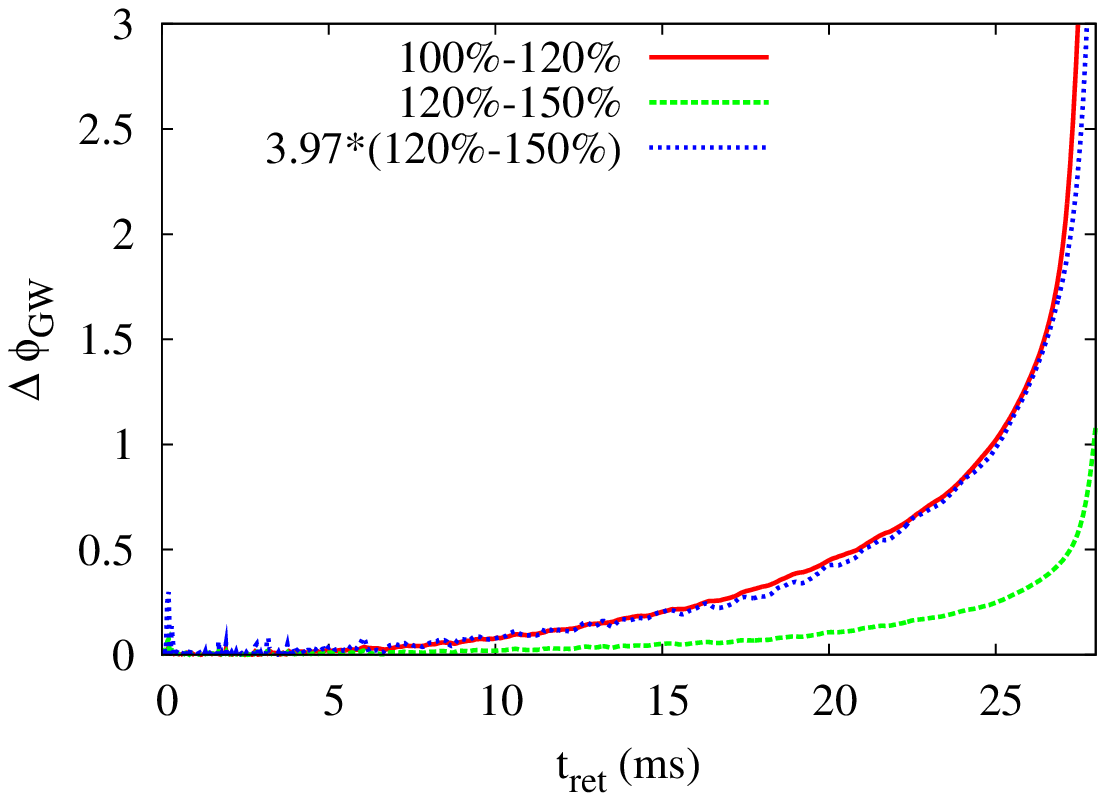}
 \end{tabular}
 \caption{Convergence property of $\Psi_{4,22}$ for H4-135-QC with
 different grid resolutions. The high, middle, and low (baseline of this
 study) resolutions are denoted by 150\%, 120\%, and 100\%,
 respectively. The left, middle, and right panels show the amplitude
 evolution, difference of amplitude evolution, and difference of phase
 evolution, respectively. We do not show the phase evolution itself,
 because differences among different grid resolutions are barely
 distinguished from the direct comparison. We also include a rescaled
 difference between high and middle resolutions using a scaling factor
 3.97 obtained assuming a hypothetical eighth-order convergence for
 difference plots.} \label{fig:convH4QC}
\end{figure*}

\begin{figure*}
 \begin{tabular}{ccc}
  \includegraphics[width=59mm,clip]{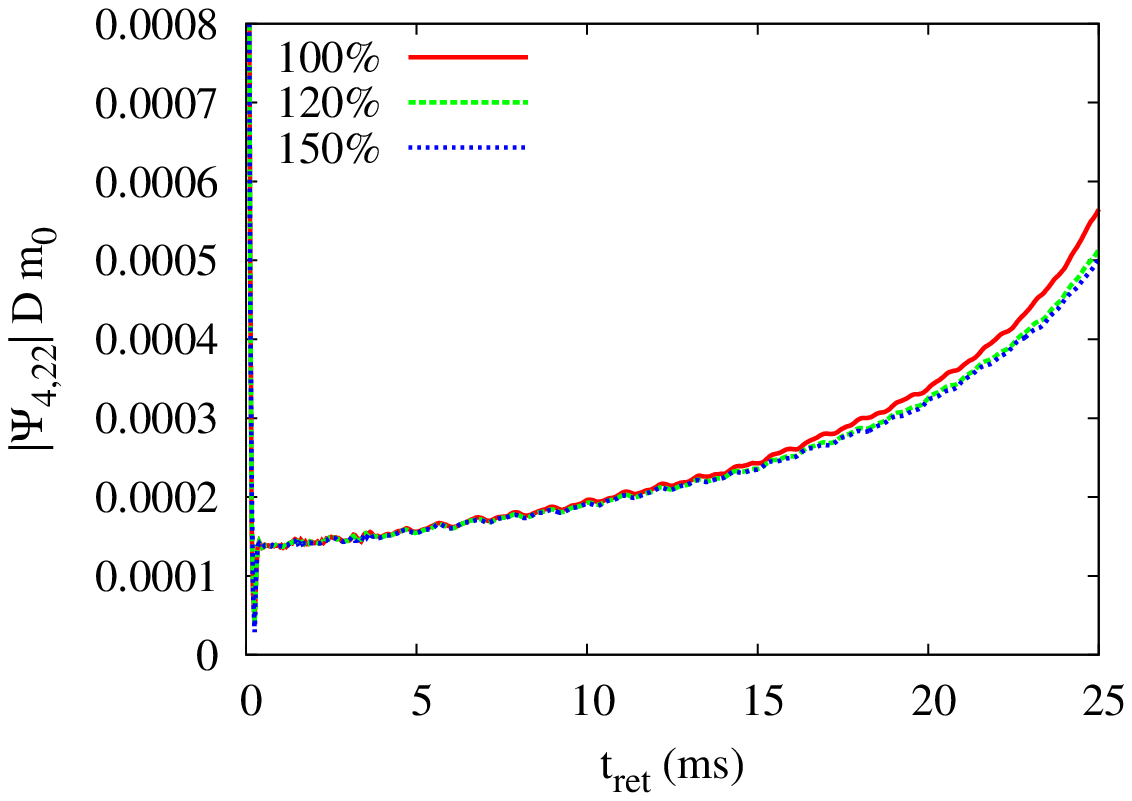} &
  \includegraphics[width=59mm,clip]{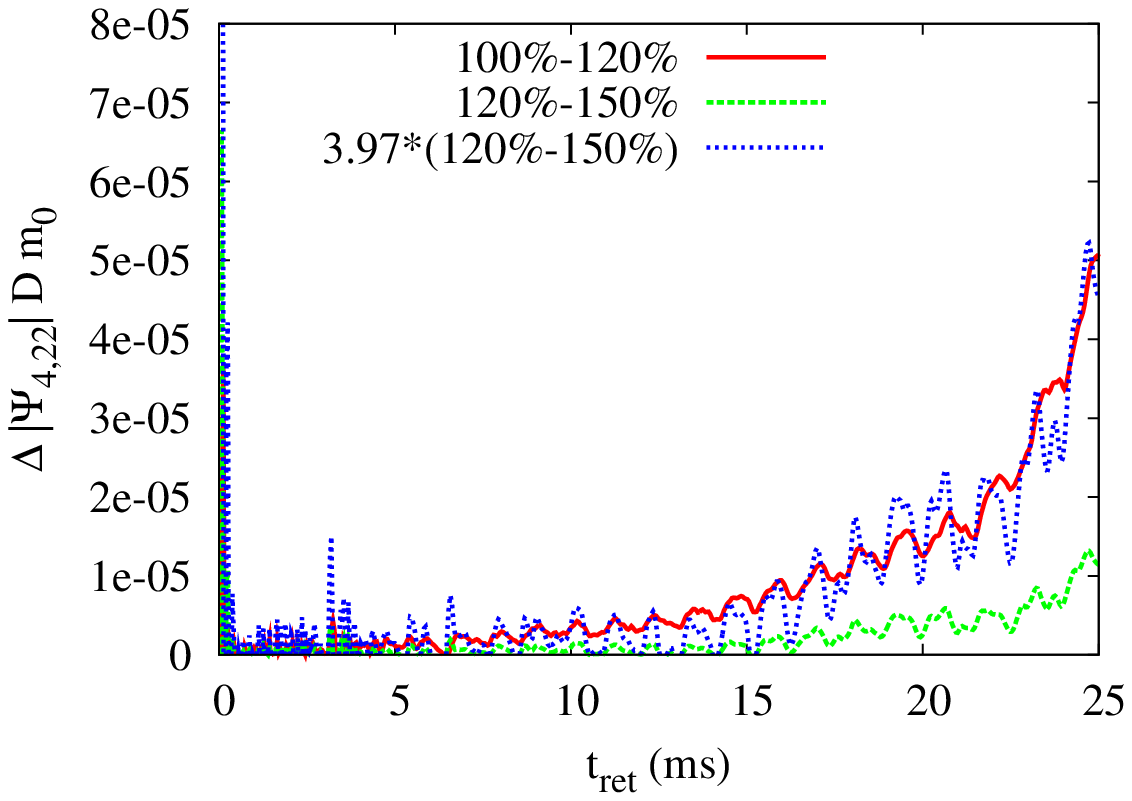} &
  \includegraphics[width=59mm,clip]{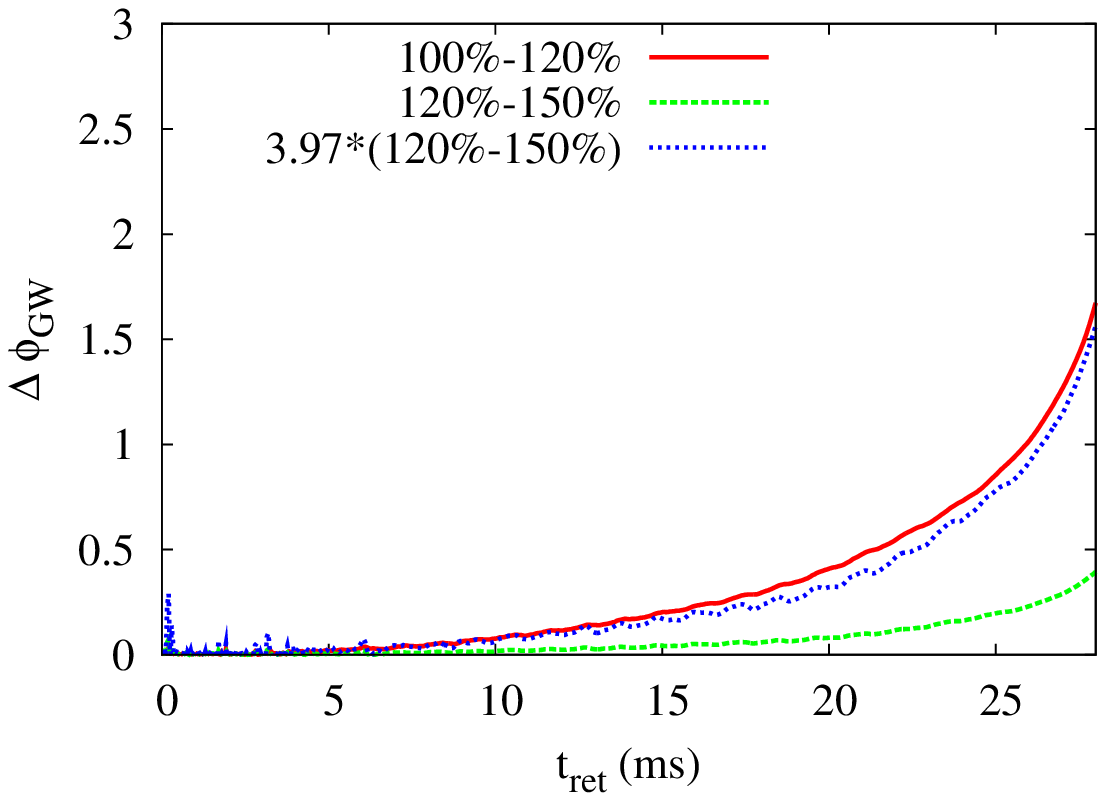}
 \end{tabular}
 \caption{The same as Fig.~\ref{fig:convH4QC}, but for H4-135-Iter3. The
 time intervals are taken to be the same as those of
 Fig.~\ref{fig:convH4QC} in all the plots. Because the merger time is
 later for H4-135-Iter3 than for H4-135-QC, the amplitude and phase
 differences at the end of the plots are smaller accordingly.}
 \label{fig:convH4fit3}
\end{figure*}

We discuss convergence issues to show that the presence and/or absence
of eccentricity-driven modulations found in various quantities are
inherent in initial data rather than associated with the finite grid
resolution. Figure \ref{fig:convH4QC} compares the evolution of
quantities associated with $\Psi_{4,22}$ for H4-135-QC among high,
middle, and low resolutions, where the low resolution has always been
adopted to derive all the results presented so far. The left panel shows
the amplitude evolution and indicates that the modulations with a $\sim$
5 ms period are in phase among all the grid resolutions. Furthermore,
the differences of the amplitude plotted in the middle panel show no
modulation with this period. These facts imply that the eccentricity is
a genuine property of initial data. Figure \ref{fig:convH4fit3} shows
the same quantities as Fig.~\ref{fig:convH4QC} but for H4-135-Iter3, and
no appreciable eccentricity-driven modulation is observed in either the
amplitude itself or its difference. This fact implies that the low
eccentricity of H4-135-Iter3 is not an artifact of a particular grid
setting and is again a genuine property of initial data.

The middle panels of Figs.~\ref{fig:convH4QC} and \ref{fig:convH4fit3}
also suggest that the result of new {\small SACRA} with the Z4c
formulation converges more rapidly than that of the previous version
\cite{yamamoto_st2008}. Although our evolution scheme is fourth order at
best in both time and space, the overall behavior of the amplitude
evolution among different resolutions implies that our results are
scaled with eighth-order convergence even right before the merger
(compare with Fig.~\ref{fig:gw}). Specifically, we assume that a
quantity $Q$ depends on the grid separation $\Delta x$ as
\begin{equation}
 Q (\Delta x) = Q_0 + E ( \Delta x )^p ,
\end{equation}
where $Q_0 \equiv Q(0)$ is the value at the continuum limit, $E$ is a
constant, and $p$ is the convergence order. We readily derive
\begin{equation}
 \frac{Q ( \Delta x ) - Q ( \Delta x / 1.2 )}{Q ( \Delta x / 1.2 ) - Q (
  \Delta x / 1.5 )} = \frac{1 - 1/(1.2)^p}{1/(1.2)^p - 1/(1.5)^p} .
\end{equation}
Our numerical results suggest that the left-hand side is approximately
equal to 4, and this is explained if $p \approx 8$. We never expect that
our new code converges with the eighth order, which might be achieved
when lower-order truncation errors happen to be very small. We guess
rather that the range of grid resolutions spanned in this study, 150\%,
is not sufficient to clarify the convergence property in a nearly
convergent regime. A more systematic investigation spanning a wide range
of grid resolutions is evidently required. It should be cautioned that
assessment of convergence properties in numerical relativity is a highly
nontrivial task, especially when a sophisticated
adaptive-mesh-refinement algorithm is adopted \cite{zlochower_pl2012}.
As {\small SACRA} has been verified in many places with the BSSN
formulation (see, e.g., Ref.~\cite{read_bcfgkmrst2013}), our results
convince us that the modulation is not an artifact of finite grid
resolutions.

The right panels of Figs.~\ref{fig:convH4QC} and \ref{fig:convH4fit3}
compare gravitational-wave phase evolution among different grid
resolutions. Again, we see no modulation with the $\sim$ 5 ms period in
the phase differences, and thus the eccentricity is shown to be inherent
in initial data. An apparent eighth-order convergence is also found in
the phase evolution in a manner consistent with the amplitude
evolution. A small phase difference of $\sim$ 1 radian between middle
and high resolutions at $t_\mathrm{ret}=28$ ms may be more important for
practical purposes, where this approximately corresponds to $m_0
\Omega_\mathrm{GW} = 0.1$ for H4-135-QC. The small difference should be
contrasted with previous results obtained in the BSSN formulation, with
which the number of orbits can differ by a factor of order unity with a
similar grid resolution \cite{hotokezaka_ks2013}. While a sophisticated
extrapolation method such as the one developed in
Ref.~\cite{hotokezaka_ks2013} will be still useful to compute physical
waveforms, slight improvement of the grid resolution would give us a
phase error smaller than 0.1 radian, accurate enough for the next
generation of interferometric detectors.

\section{Summary and discussion} \label{sec:summary}

We developed a method to obtain low-eccentricity initial data of binary
neutron stars for numerical relativity. In the beginning, we computed
standard quasicircular initial data assuming helical symmetry with the
angular velocity determined by force balance at the center of neutron
stars
\cite{gourgoulhon_gtmb2001,taniguchi_gourgoulhon2002,taniguchi_gourgoulhon2003,taniguchi_shibata2010},
and evolved them for $\sim$ 3 orbits. We fit the time derivative of the
coordinate orbital angular velocity to an analytic function and
estimated appropriate corrections to the orbital angular velocity and
approaching velocity for the eccentricity reduction. We then modified
the initial data by adjusting the orbital angular velocity and
approaching velocity. This modification affected the solution primarily
through the hydrostatic equilibrium equations for binary neutron stars
and should be contrasted with the case of binary black holes
\cite{pfeiffer_bklls2007,husa_hgsb2007}. We repeated these procedures
(initial data computation, time evolution, and analysis of the orbit)
until the eccentricity was reduced to a desired value.

We demonstrated the ability of our eccentricity reduction by simulating
two families of equal-mass binary neutron stars. The eccentricity was
decreased from $\sim$ 0.01 to $\lesssim$ 0.001 by three successive
iterations beginning with quasicircular initial data. We found that
low-eccentricity initial data exhibit smaller modulations in evolution
of the orbital separation, gravitational-wave amplitude, and
gravitational-wave frequency than quasicircular initial data. Smooth
evolution of gravitational-wave quantities associated with
low-eccentricity initial data will help comparisons with analytic models
and hybridization to construct theoretical templates. We also found that
the accuracy of gravitational waves derived by low-eccentricity initial
data is limited by high-frequency oscillations, which might be ascribed
to insufficient outer boundary conditions of {\small SACRA}.

Aside from gravitational waves investigated in this study, the
eccentricity could affect properties of remnant massive neutron stars
and mass ejection from the system, and hence electromagnetic
signals. Indeed, simulations performed in this study suggest that the
ejecta mass, which is $3 \times 10^{-4} M_\odot$ for H4-135-QC and $0.02
M_\odot$ for APR4-14-QC at 10 ms after the merger, seems to vary by
$O(10\%)$ between quasicircular and low-eccentricity initial data. This
variation is not always systematic with respect to the eccentricity,
possibly because the velocity at the contact of binary neutron stars can
fluctuate within the eccentricity-driven modulation in each particular
simulation. We do not present results for them in detail here, because
quantitative conclusions require systematic investigations with
convergence analysis. Remnant massive neutron stars and mass ejection
will be relevant to short-hard gamma-ray bursts (see
Refs.~\cite{nakar2007,berger2014} and references therein for reviews)
and other electromagnetic counterparts
\cite{li_paczynski1998,nakar_piran2011}, and thus the eccentricity
reduction may also be important to quantitatively clarify
electromagnetic signals from binary neutron star mergers.

\begin{acknowledgments}
 Koutarou Kyutoku is grateful to John L. Friedman for careful reading of
 the manuscript and Hiroyuki Nakano for valuable discussions. This work
 is supported by Grant-in-Aid for Scientific Research (Grant
 No.~24244028), and Koutarou Kyutoku is supported by JSPS Postdoctoral
 Fellowship for Research Abroad.
\end{acknowledgments}

\appendix

\section{Post-Newtonian formula} \label{app:pn}

The energy and angular momentum of a binary of nonspinning masses $m_1$
and $m_2$ with the angular velocity $\Omega$ are currently known up to
fourth post-Newtonian order for point-particle contributions
\cite{blanchet2014}. Here, the total mass is defined by $m_0 \equiv m_1
+ m_2$, and the symmetric mass ratio is defined by $\nu \equiv m_1 m_2 /
m_0^2$. A post-Newtonian parameter is defined by
\begin{equation}
 x \equiv \left( \frac{G m_0 \Omega}{c^3} \right)^{2/3} ,
\end{equation}
where $G$ and $c$ are inserted for clarity in this Appendix. The orbital
binding energy is given by
\begin{widetext}
 \begin{align}
  \frac{E}{m_0 c^2} = - \frac{\nu x}{2} \biggl\{ & 1 + \left( -
  \frac{3}{4} - \frac{\nu}{12} \right) x + \left( - \frac{27}{8} +
  \frac{19}{8} \nu - \frac{\nu^2}{24} \right) x^2 \notag \\
  + & \left( - \frac{675}{64} + \left[ \frac{34445}{576} -
  \frac{205}{96} \pi^2 \right] \nu - \frac{155}{96} \nu^2 -
  \frac{35}{5184} \nu^3 \right) x^3 \notag \\
  + & \biggl( - \frac{3969}{128} + \left[ - \frac{123671}{5760} +
  \frac{9037}{1536} \pi^2 + \frac{1792}{15} \ln 2 + \frac{896}{15}
  \gamma_\mathrm{E} \right] \nu \notag \\
  + & \left[ - \frac{498449}{3456} + \frac{3157}{576} \pi^2 \right]
  \nu^2 + \frac{301}{1728} \nu^3 + \frac{77}{31104} \nu^4 +
  \frac{448}{15} \nu \ln x \biggr) x^4 \biggr\} ,
 \end{align}
 where $\gamma_\mathrm{E}$ is Euler's constant. The orbital angular
 momentum is given by
 \begin{align}
  \frac{J}{G m_0^2 /c} = \frac{\nu}{x^{1/2}} \biggl\{ & 1 + \left(
  \frac{3}{2} + \frac{\nu}{6} \right) x + \left( \frac{27}{8} -
  \frac{19}{8} \nu + \frac{\nu^2}{24} \right) x^2 \notag \\
  + & \left( \frac{135}{16} + \left[ - \frac{6889}{144} + \frac{41}{24}
  \pi^2 \right] \nu + \frac{31}{24} \nu^2 + \frac{7}{1296} \nu^3 \right)
  x^3 \notag \\
  + & \biggl( \frac{2835}{128} + \left[ \frac{98869}{5760} -
  \frac{6455}{1536} \pi^2 - \frac{256}{3} \ln 2 - \frac{128}{3}
  \gamma_\mathrm{E} \right] \nu \notag \\
  + & \left[ \frac{356035}{3456} - \frac{2255}{576} \pi^2 \right] \nu^2
  - \frac{215}{1728} \nu^3 - \frac{55}{31104} \nu^4 - \frac{64}{3} \nu
  \ln x \biggr) x^4 \biggr\} .
 \end{align}
\end{widetext}
The latter is derived from the former using a so-called thermodynamic
relation,
\begin{equation}
 \frac{\partial E}{\partial \Omega} = \Omega \frac{\partial J}{\partial
  \Omega} ,
\end{equation}
or equivalently
\begin{equation}
 \frac{\partial [ E / ( m_0 c^2 ) ]}{\partial x} = x^{3/2}
  \frac{\partial [ J / (G m_0^2 / c) ]}{\partial x} .
\end{equation}

Finite-size contributions are computed up to first post-Newtonian order
to linear quadrupolar tidal deformation \cite{vines_flanagan2013}. We
parametrize the finite-size effect of a neutron star with the
gravitational mass $m$ and radius $R$ by a dimensionless quadrupolar
tidal deformability defined by \cite{lackey_ksbf2012,lackey_ksbf2014}
\begin{equation}
 \Lambda = G \lambda \left( \frac{c^2}{G m} \right)^5 = \frac{2}{3} k
  \left( \frac{c^2 R}{G m} \right)^5 ,
\end{equation}
where $\lambda$ and $k$ are the quadrupolar tidal deformability and Love
number, respectively. Specific values of $\Lambda$ are 1111 for a $1.35
M_\odot$ neutron star with the H4 equation of state and 256 for a $1.4
M_\odot$ neutron star with the APR4 equation of state. Hereafter, the
dimensionless deformability of $m_1$ and $m_2$ are denoted by
$\Lambda_1$ and $\Lambda_2$, respectively. We further define $q_1 \equiv
m_1 / m_0$ and $q_2 \equiv m_2 / m_0$. The contribution to the orbital
binding energy is
\begin{align}
 \frac{E_\mathrm{tidal}}{m_0 c^2} = \frac{\nu x}{2} \biggl\{ 9 \biggl( &
 q_1^4 q_2 \Lambda_1 + q_1 q_2^4 \Lambda_2 \biggr) x^5 \notag \\
 + \frac{11}{2} \biggl[ & ( 3 + 2 q_1 + 3 q_1^2) q_1^4 q_2 \Lambda_1
 \notag \\
 + & ( 3 + 2 q_2 + 3 q_2^2 ) q_1 q_2^4 \Lambda_2 \biggr] x^6 \biggr\} ,
\end{align}
and that to the orbital angular momentum is
\begin{align}
 \frac{J_\mathrm{tidal}}{G m_0^2 / c} = \frac{\nu}{x^{1/2}} \biggl\{ 6
 \biggl( & q_1^4 q_2 \Lambda_1 + q_1 q_2^4 \Lambda_2 \biggr) x^5
 \notag \\
 + \frac{7}{2} \biggl[ & ( 3 + 2 q_1 + 3 q_1^2 ) q_1^4 q_2 \Lambda_1
 \notag \\
 + & ( 3 + 2 q_2 + 3 q_2^2 ) q_1 q_2^4 \Lambda_2 \biggr] x^6 \biggr\},
\end{align}
where the thermodynamic relation is used again. These contributions are
simply added to the point-particle contributions described above. For $G
m_0 \Omega / c^3 = 0.019$ considered in this study, point-particle terms
up to second post-Newtonian order dominate the energy and angular
momentum, while the sum of higher-order terms and finite-size
corrections contribute only by $\sim$ 0.1\%--0.2\% even for a relatively
stiff H4 equation of state.

\section{Computation of gravitational waves} \label{app:gw}

In this Appendix, we summarize our derivation of gravitational waveforms
$h = h_+ - \mathrm{i} h_\times$ from $\Psi_4$ obtained by numerical
simulations. We first extrapolate $\Psi_4$ extracted at a finite
coordinate radius, $r_\mathrm{ex}$, to null infinity following
Ref.~\cite{lousto_nzc2010}. Using the areal radius defined by
Eq.~\eqref{eq:areal}, we compute
\begin{align}
 & \left. D \Psi_{4,\ell m} (t) \right|_{D \to \infty} = \left( 1 -
 \frac{2m_0}{D} \right) \notag \\
 & \times \left[ D \Psi_{4,\ell m} (t) - \frac{( \ell - 1 ) ( \ell + 2
 )}{2} \int \Psi_{4, \ell m} (t') dt' \right] \label{eq:extrapolate} ,
\end{align}
where $\Psi_{4,\ell m}$ is the $( \ell , m )$-mode coefficient of
$\Psi_4$ projected onto spin-weighted spherical harmonics. The prefactor
$1 - 2m_0/D$ approximately corrects a difference between the tetrad used
in {\small SACRA} \cite{yamamoto_st2008} and the Kinnersly tetrad, where
the latter has to be chosen to derive this extrapolation formula. Next,
gravitational waveforms are computed by integrating this extrapolated
$\Psi_4$ (or $D \Psi_4$) twice as
\begin{equation}
 h_{\ell m} (t) = \int \left( \int \Psi_{4,\ell m} (t'') dt''
			      \right) dt' .
\end{equation}
The angular frequency of gravitational waves, $\Omega_\mathrm{GW} (t)$,
is estimated by
\begin{equation}
 \Omega_{\mathrm{GW},\ell m} (t) \equiv \frac{| \int \Psi_{4,\ell m}
  (t') dt'|}{| \int \left( \int \Psi_{4,\ell m} (t'') dt'' \right) dt'
  |} .
\end{equation}
In the body text, we suppress the subscript $\ell m$ except for
$\Psi_{4,22}$, because we focus only on the $(2,2)$ mode.

All the time integrations are performed by fixed-frequency integration
\cite{reisswig_pollney2011}. The time-domain data are transformed to
frequency-domain data by
\begin{equation}
 \tilde{\Psi}_{4,\ell m} (f) = \int w (t) \Psi_{4,\ell m} (t) e^{-2 \pi
  \mathrm{i} f t} dt , \label{eq:fourier}
\end{equation}
where we apply a tapered-cosine filter of the form
\begin{equation}
 w (t) =
  \begin{cases}
   \{ 1 - \cos [ \pi ( t - t_i ) / \Delta t ] \} / 2 & ( t_i \le t <
   t_{i+} ) \\
   1 & ( t_{i+} \le t < t_{f-} ) \\
   \{ 1 - \cos [ \pi ( t_f - t ) / \Delta t ] \} / 2 & ( t_{f-} \le t <
   t_f )
  \end{cases}
  . \label{eq:taper}
\end{equation}
Here, $t_i$ and $t_f$ are the initial and final times of the data,
respectively, and $t_{i+} \equiv t_i + \Delta t$ and $t_{f-} \equiv t_f
- \Delta t$ are determined by a width of the tapering region $\Delta
t$. We choose $\Delta t \approx$ 1 ms for the computation of waveforms.
Gravitational waveforms are computed as
\begin{equation}
 h_{\ell m} (t) = - \int \frac{\tilde{\Psi}_{4,\ell m} (f)}{( 2 \pi
  \max[f,f_0] )^2} e^{2 \pi \mathrm{i} f t} df ,
\end{equation}
replacing the time integration with multiplication by $- \mathrm{i}/ ( 2
\pi f )$. Here, a fixed frequency $f_0$ is introduced to suppress
unphysical drifts associated with spurious contributions from
low-frequency components, and we choose it to be $f_0 = 0.8 m ( \Omega /
2 \pi )$ so that physical information of gravitational waves is not
affected. This integration method is also used in evaluating the
right-hand side of Eq.~\eqref{eq:extrapolate}.

Gravitational-wave spectra are computed by Eq.~\eqref{eq:fourier} and
double multiplication of $- \mathrm{i} / ( 2 \pi f)$. For this purpose,
we apply the tapered-cosine filter, Eq.~\eqref{eq:taper}, with $\Delta t
\approx$ 5 ms to remove unphysical noises (in the frequency domain)
associated with junk radiation and finite data length.

\end{document}